\documentclass[iop]{emulateapj}

\def\fnm{iPTF\,16fnm}
\def\ha{SN~2008ha}
\def\kms{$\mathrm{km \; s}^{-1}$}

\usepackage[colorlinks,urlcolor=blue,citecolor=blue,linkcolor=blue]{hyperref}
\usepackage{booktabs}
\usepackage{natbib,amsmath,textcomp}
\usepackage{pifont} 
\slugcomment{DRAFT \today}
\shorttitle{\fnm: an 02cx-like SN}
\shortauthors{Miller et al.}

\begin{document}

\title{Color Me Intrigued: \\
the Discovery of \fnm, a Supernova 2002cx-like Object}

\author{A.~A.~Miller\altaffilmark{1,2,3*}, 
M.~M.~Kasliwal\altaffilmark{3},
Y.~Cao\altaffilmark{4}, \\
A.~Goobar\altaffilmark{5},
S.~Kne\v{z}evi\'{c}\altaffilmark{6},
R.~R.~Laher\altaffilmark{7},
R.~Lunnan\altaffilmark{3},
F.~J.~Masci\altaffilmark{7},
P.~E.~Nugent\altaffilmark{8,9},
D.~A.~Perley\altaffilmark{10,11},
T.~Petrushevska\altaffilmark{5},
R.~M.~Quimby\altaffilmark{12},
U.~D.~Rebbapragada\altaffilmark{13},
J.~Sollerman\altaffilmark{14},
F.~Taddia\altaffilmark{14},
\& S. R. Kulkarni\altaffilmark{3}
}

\altaffiltext{1}{Center for Interdisciplinary Exploration and Research in Astrophysics (CIERA) and Department of Physics and Astronomy, Northwestern University, 2145 Sheridan Road, Evanston, IL 60208, USA}
\altaffiltext{2}{The Adler Planetarium, Chicago, IL 60605, USA}
\altaffiltext{3}{Division of Physics, Mathematics, and Astronomy, California Institute of Technology, Pasadena, CA 91125, USA}
\altaffiltext{4}{eScience Institute and Astronomy Department, University of Washington, Seattle, WA 98195, USA}
\altaffiltext{5}{The Oskar Klein Centre, Department of Physics, AlbaNova, SE-106 91 Stockholm, Sweden}
\altaffiltext{6}{Department of Particle Physics and Astrophysics, Weizmann Institute of Science, Rehovot 7610001, Israel}
\altaffiltext{7}{Infrared Processing and Analysis Center, California Institute of Technology, Pasadena, CA 91125, USA}
\altaffiltext{8}{Lawrence Berkeley National Laboratory, Berkeley, California 94720, USA}
\altaffiltext{9}{University of California -- Berkeley, Berkeley, CA 94720, USA}
\altaffiltext{10}{Dark Cosmology Centre, Niels Bohr Institute, University of Copenhagen, Juliane Maries Vej 30, DK-2100 K{\o}benhavn {\O}, Denmark}
\altaffiltext{11}{Liverpool John Moores University, Department of Astronomy, Liverpool, L3 5RF, UK}
\altaffiltext{12}{Department of Astronomy, San Diego State University, San Diego, CA 92182, USA}
\altaffiltext{13}{Jet Propulsion Laboratory, California Institute of Technology, Pasadena, CA 91109, USA}
\altaffiltext{14}{The Oskar Klein Centre, Department of Astronomy, AlbaNova, SE-106 91 Stockholm, Sweden}
\altaffiltext{*}{E-mail: {\tt amiller@northwestern.edu}}

\begin{abstract}

Modern wide-field, optical time-domain surveys must solve a basic optimization problem: maximize the number of transient discoveries or minimize the follow-up needed for the new discoveries. Here, we describe the \textit{Color Me Intrigued} experiment, the first from the intermediate Palomar Transient Factory (iPTF) to search for transients simultaneously in the $g_\mathrm{PTF}$- and $R_\mathrm{PTF}$-bands. During the course of this experiment we discovered \fnm, a new member of the 02cx-like subclass of type Ia supernovae (SNe). \fnm\ peaked at $M_{g_\mathrm{PTF}} = -15.09 \pm 0.17 \; \mathrm{mag}$, making it the second least-luminous known type Ia SN. \fnm\ exhibits all the hallmarks of the 02cx-like class: (i) low luminosity at peak, (ii) low ejecta velocities, and (iii) a non-nebular spectra several months after peak. Spectroscopically, \fnm\ exhibits a striking resemblence to 2 other low-luminosity 02cx-like SNe: SNe~2007qd and 2010ae. \fnm\ and SN~2005hk decline at nearly the same rate, despite a 3 mag difference in brightness at peak. When considering the full subclass of 02cx-like SNe, we do not find evidence for a tight correlation between peak luminosity and decline rate in either the $g'$ or $r'$ band. We further examine the $g' - r'$ evolution of 02cx-like SNe and find that their unique color evolution can be used to separate them from 91bg-like and normal type Ia SNe. This selection function will be especially important in the spectroscopically incomplete Zwicky Transient Facility/Large Synoptic Survey Telescope era. We measure the relative rate of 02cx-like SNe to normal SNe Ia and find $r_{N_{02cx}/N_{Ia}} = 25^{+75}_{-18.5}\%$. Finally, we close by recommending that LSST periodically evaluate, and possibly update, its observing cadence to maximize transient science.

\end{abstract}


\section{Introduction}\label{intro}

The proliferation of large-area optical detectors has led to a recent renaissance of time-domain astronomy, and, in particular, the search for exotic transients. Over the past $\sim$decade, new surveys have dramatically increased the number of transients discovered on a nightly basis, and these efforts will culminate with the Large Synoptic Survey Telescope (LSST) in the early 2020s, which will discover 2,000 new supernovae (SNe) per night \citep{Ivezic08}. Discovery is a small first step in improving our understanding of SNe. Detailed follow-up observations, either photometry spanning the ultraviolet, optical and infrared (UVOIR) or spectra, are needed to reveal the nature of these explosions (see \citealt{Filippenko97} for a review of SNe spectra). Existing follow-up facilities are already taxed by the current rate of transient discovery, meaning the LSST-enabled two orders of magnitude increase in the discovery rate poses a serious ``follow-up problem.'' Namely, there will be substantially more known SNe than available resources to study them all in detail. 

All transient surveys are forced to make a decision regarding the trade-off between discovery potential and the need for follow-up resources. For example, to maximize the number of transient discoveries, a survey should observe as wide an area as possible in a single filter. Implicit in this strategy is the need for outside follow-up resources. If, on the other hand, follow-up resources are scarce or prohibitively expensive, a survey may choose to repeatedly observe the same fields in different passbands to obtain color information. This would, however, reduce the total number of transient discoveries. Both strategies are employed by modern surveys. Broadly speaking, shallow surveys tend to observe in a single filter [e.g., the All-Sky Automated Survey for Supernova (ASAS-SN; \citealt{Shappee14}); the Palomar Transient Factory (PTF; \citealt{law09})], while deeper surveys sacrifice area for color [e.g., the Supernova Legacy Survey (SNLS; \citealt{Astier06}); the Panoramic Survey Telescope and Rapid Response System 1 (PS1) Medium Deep Survey (MDS; \citealt{Chambers16}), the Dark Energy Survey Supernova search (DES SN survey; \citealt{Kessler15})]. The looming LSST ``follow-up problem'' has led to an increasing number of recent studies that consider only the photometric evolution of transients (e.g., \citealt{Jones16}).

The intermediate Palomar Transient Factory (iPTF; \citealt{Kulkarni13}), which succeeded PTF,\footnote{The initial PTF survey was conducted from 2009 Jul.\ through 2012 Dec. The iPTF survey was conducted from 2013 Feb.\ through 2016 Oct. Finally, the iPTF-extension was conducted from 2016 Nov.\ through 2017 Feb.} is a time-domain survey that has been conducted as a series of experiments. Like its predecessor, iPTF searched for transients using a single filter while obtaining two observations per field per night to reject asteroids from the transient candidate stream. Here, we describe the \textit{Color Me Intrigued} experiment from the final semester of iPTF. This experiment was the first from PTF/iPTF to search for transients with multiple filters, though this color information was achieved with no loss of efficiency as the experiment still obtained two observations per field per night. 

To our knowledge, \textit{Color Me Intrigued} is the first transient experiment to employ both near-simultaneous color observations, $g_\mathrm{PTF}$ and $R_\mathrm{PTF}$ in this case, and a 1-d search cadence. Many high-$z$ type Ia SN surveys have obtained near-simultaneous colors (often in as many as 5 filters), though the revisit time is always greater than 1 d, and typically $\sim$3--5 d (e.g., SNLS, \citealt{Astier06}; the Sloan Digital Sky Survey-II Supernova Survey, \citealt{Frieman08}; DES SN survey, \citealt{Kessler15}). The ASAS-SN survey searches for transients with a 2-d cadence, but all ASAS-SN observations are obtained in a single broadband filter \citealt{Shappee14}. The closest comparison to the \textit{Color Me Intrigued} search strategy was that employed by the PS1 MDS. PS1 MDS obtained images of 10 separate $\sim$7 $\deg^2$ fields, every night a given field was visible. During dark time, observations would be obtained in both the $g'$ and $r'$ filters on the same night with a 3-d cadence \citep{Chambers16}. On the first and second nights following $g'r'$ observations, PS1 MDS would observe exclusively in the $i'$ and $z'$ filters, respectively. Thus, while PS1 MDS had a 1-d cadence, near-simultaneous colors were obtained at best every 3 nights.

During the course of the \textit{Color Me Intrigued} experiment, iPTF discovered a rare SN~2002cx-like object (hereafter 02cx-like SN), \fnm. We discuss this SN in detail below.

\subsection{Peculiar 02cx-like SNe}

The discovery of an accelerating universe \citep{Riess98,Perlmutter99} has led to numerous and extensive observational and theoretical studies into the nature of type Ia SNe over the past two decades. Despite these efforts, a precise understanding of the nature and exact explosion mechanium for SN Ia progenitor systems is still unknown. While there is strong observational evidence that at least some SNe Ia come from white dwarf (WD) systems (e.g., \citealt{Bloom12a}), and a general consensus that carbon/oxygen WDs give rise to type Ia SNe, there is ambiguity in the mechanisms and scenarios that lead to explosion (e.g., \citealt{Hillebrandt13}). The multitude of observational campaigns designed to capture large samples of type Ia SNe for cosmological studies have also revealed several peculiar hyrodgen-poor SNe (see \citealt{Kasliwal12a} and references therein). While these peculiar SNe retain many similarities to normal type Ia SNe, their distinct properties allow for the possibility of more extreme or unusual WD progenitor scenarios.

To date, the most common sub-class of peculiar SNe Ia are those similar to SN~2002cx \citep{Li03}, of which there are now $\sim$2 dozen known examples (e.g., \citealt{Foley13}). The 02cx-like\footnote{This subclass is sometimes alternatively referred to as type Iax SNe \citep{Foley13}.} class is characterized by low ejecta velocities ($\sim$half normal SNe Ia) and low luminosities, ranging from $M \approx -19$ to $-14$~mag at peak. The distribution of host-galaxy morphologies for 02cx-like SNe is strongly skewed towards late-type hosts (e.g., \citealt{Foley13,White15} and references therein), which may indicate massive star progenitors for this class (e.g., \citealt{Valenti09,Moriya10}). However, the maximum-light spectrum of SN~2008ha shows clear evidence for carbon/oxygen burning, providing a strong link between 02cx-like SNe and WD progenitors \citep{Foley10}.

While the sample of 02cx-like SNe is relatively small, they constitute a signifigant fraction of the total number of type Ia SNe, $\sim$5--30\% \citep{Li11,Foley13,White15}. Multiple efforts have been made to identify simple correlations between basic observational properties for the class (e.g., \citealt{McClelland10,Foley13}), in part to aid the identification of a likely progenitor scenario. Significant outliers exist (e.g., \citealt{Narayan11}), however, and the emerging consensus seems to be that 02cx-like SNe cannot be understood as a single parameter family \citep{Magee17}. 

Pure deflagration models are often invoked to explain the heterogenity of the 02cx-like class (e.g., \citealt{Phillips07}), as they can naturally explain the wide range in peak luminosity. As a result, extensive theoretical consideration has been given to these models recently (e.g., \citealt{Kromer13,Fink14}), with the express purpose of understanding 02cx-like SNe. Recently, \citet{Magee16} compared early spectroscopic observations of SN~2015H, a member of the 02cx-class, to a deflagration model from \citet{Fink14}, and found good agreement. The model light curve evolves faster than the observations, though this may be reconciled with higher-ejecta-mass models \citep{Magee16}. An interesting consequence of the pure deflagration models is that they do not completely unbind the WD, leaving a $\sim$1 $M_\odot$ bound remnant \citep{Kromer13}. Alternatively, \citet{Stritzinger15} compared detailed observations of SN~2012Z, one of the most luminous members of the 02cx-like class, to pulsating delayed-detonation (PDD) models of exploding white dwarfs. In particular, \citeauthor{Stritzinger15} observe pot-bellied [\ion{Fe}{2}] profiles in the near-infrared at late times, which are an indication of high-density burning. Based on this observation, and the direct comparison of model light curves and spectra, it is argued that SN~2012Z was a PDD explosion of a Chandrasekhar mass WD and not a pure deflagration \citep{Stritzinger15}. While examining a large sample of low-velocity type I SNe, it is argued in \citet{White15} that 02cx-like SNe are the result of double degenerate mergers, which may explain their observed heterogenity as the masses of merging WDs can vary significantly. \citeauthor{White15} prefer pure deflagration models as an explanation for 2002es-like SNe, a low-velocity subclass of type Ia SNe that is distinct from the 02cx-like class \citep{Ganeshalingam12,Cao16a}. Moving forward, it is clear that additional members of the 02cx-like class need to be found to disambiguate between pure deflagrations, PDD explosions, double degenerate mergers, and alternative scenarios.

Here, we present the discovery of a new member of the 02cx-like class, \fnm. \fnm\ was discovered in the course of the iPTF \textit{Color Me Intrigued} experiment, and it was first identified as a possible low-velocity SN with spectra from the Spectral Energy Distribution Machine\footnote{Documentation for SEDm is available here: \url{http://www.astro.caltech.edu/sedm/}.} (SEDm; Blagoradnova et al.\ 2017, in prep.). Our spectroscopic follow-up campaign clearly identfies \fnm\ as an 02cx-like SN, which we show is one of the faintest known members of the class. The $g_\mathrm{PTF}$ and $R_\mathrm{PTF}$ light curves of \fnm\ allow us to compare it, and several other 02cx-like SNe, to both normal and underluminous type Ia SNe.

\section{Color Me Intrigued: The iPTF Two Filter Experiment}

The iPTF survey has been organized as a series of time-domain experiments conducted with the Palomar 48-in telescope (P48) selected on a semesterly basis. Proposals for individual experiments are written by members of iPTF institutions and selected by an internal time allocation committee. The decision to focus on a single experiment at a time was to minimize cadence interruptions and ensure a specific science goal could be accomplished. While the precise details (e.g., survey area, cadence, etc.) of the individual experiments varied, seasonal weather patterns and target availability constrained the possible science returns. As a result, experiments rotated on a roughly quarterly schedule with the following aims:\footnote{Again, this is a broad overview which does not provide exact details of each experiment.} (i) winter -- persistent, but unpredictable, P48 weather closures led to wide-area surveys searching for intranight variability (e.g., \citealt{Cenko13}) and/or rare, slowly evolving events (such as superluminous SNe), as it is otherwise difficult to maintain a night-to-night cadence, (ii) spring -- smaller area surveys focused on maintaining a 1-d cadence to identify young SNe (e.g., \citealt{Cao16}), (iii) summer -- wide-area surveys monitored the galactic plane (e.g., \citealt{Bellm16}), and (iv) fall -- 1-d cadence similar to the spring experiments.

Recognizing that the next generation Zwicky Transient Facility (ZTF; \citealt{Kulkarni16,Bellm16a}) would increase the discovery rate relative to PTF/iPTF by more than an order of magnitude, we (AAM, YC, \& MMK) proposed a ZTF pilot survey designed to reduce the ``follow-up problem'' for the final semester of iPTF. In brief, the proposed experiment, titled \textit{Color Me Intriguied}, would be the first by PTF/iPTF to survey simultaneously in the $g_\mathrm{PTF}$ and $R_\mathrm{PTF}$ filters. Even with the advent of fast and efficient spectroscopic resources (e.g., SEDm), ZTF transient discoveries will outpace follow-up capabilities. Adopting a two filter strategy allows us to maximize the information content for each newly discovered transient, without compromising the overall survey area. 

To reject asteroids as false positives in the search for transients, PTF/iPTF has always obtained a minimum of 2 observations of the same fields within a given night. These observations, typicaly separated by $\sim$0.5 hr, could then reject moving objects by requiring transient candidates to have at least 2 co-spatial detections separated by $\gtrsim$0.5 hr. Prior to \textit{Color Me Intrigued}, these observations were always obtained in the same filter, with the initial PTF survey designed to observe in the $g_\mathrm{PTF}$ band during dark time and the $R_\mathrm{PTF}$ band during bright time \citep{law09}. \textit{Color Me Intriguied} retained the basic 2 observations per field per night strategy, but instead observed once in $g_\mathrm{PTF}$ and once in $R_\mathrm{PTF}$, regardless of moon phase. Thus, for the first time during PTF/iPTF all newly discovered transients would have quasi-simultaneous color coverage for the duration of their evolution. This color information reduces the ``follow-up'' problem in multiple ways. First, it enables an important triage of all newly discovered candidates, as the color provides an initial classification (e.g, \citealt{Poznanski02}). Furthermore, the transients with the most extreme $g_\mathrm{PTF} - R_\mathrm{PTF}$ colors can immediately trigger follow-up observations. Second, for the subset of sources where follow-up observations prove impossible color curves greatly improve photometric classification. Finally, we note that few transients show intra-night variability, meaning a second observation in the same filter only serves the purpose of rejecting moving objects. Changing filters accomplishes this same goal, while also adding information.

This two filter strategy needed to be tested via pilot survey prior to ZTF to address one major and one minor concern. The minor concern is that using two filters would somehow prevent the rejection of asteroids as non-transients. Given that asteroids are primarily rejected via their motion, however, filter choices should have no affect on our ability to identify genuine transients. Indeed, in practice we found that maintaining the requirement of two detections in the same night rejected asteroids from the transient-candidate stream. The major concern is that young transients may be missed due to (i) extreme colors, (ii) differences in the depth of the $g_\mathrm{PTF}$ and $R_\mathrm{PTF}$ observations, or (iii) a combination of the two. For example, most young transients are hot and therefore have blue $g_\mathrm{PTF} - R_\mathrm{PTF}$ colors. If the hot transient produces a faint detection in the $g_\mathrm{PTF}$ band, then there may be no corresponding detection in the $R_\mathrm{PTF}$ band, resulting in a non-detection. In this scenario the transient would have been detected with 2 $g_\mathrm{PTF}$ band observations. Additionally, iPTF $g_\mathrm{PTF}$ band observations reach a flux limit that is a factor of $\sim$2 fainter than $R_\mathrm{PTF}$ band \citep{law09}. Thus, even relatively red sources ($g_\mathrm{PTF} - R_\mathrm{PTF} \gtrsim 0.5 \; \mathrm{mag}$) detected near the $g_\mathrm{PTF}$ band limit could be missed. A full systematic study of the biases introduced by the \textit{Color Me Intrigued} strategy is beyond the scope of this paper, and will be addressed in a future study (A.\ A.\ Miller et al. 2017, in prep.). The discovery of 1-d old SNe and rare transients (e.g., iPTF~16fnl; \citealt{Blagorodnova17}), however, suggests that the two-filter strategy does not significantly reduce the discovery potential of iPTF or ZTF. 

The \textit{Color Me Intrigued} experiment included 270 fields covering a total area of $\sim$1940 $\deg^2$. The experiment was conducted for 3 months, from 2017 Aug.\ 20 to Nov.\ 10, using the $\sim$21 darkest nights during each lunation in that timeframe.\footnote{The $\sim$7 nights centered on full moon were used to complete the iPTF Census of the Local Universe H$\alpha$ survey \citep{Cook17}.} As a ZTF/LSST precursor the experiment adopted a rolling cadence strategy. The 270 fields were split into 3 groups of 90. During each lunation, one group would be obversed with a 1-d cadence, while the other two would be observed with a 3-d cadence. Therefore each night 150 fields (90 from the 1-d cadence group and 60 from the 3-d cadence group) would be observed, yielding 300 total observations. Over the course of the experiment each field was slated to be observed 42 times, though weather losses ultimately reduced this number slightly. Adopting a rolling cadence strategy allowed us to obtain observations with a 1-d cadence, which is crucial for the study of young SNe, while maintaining a uniform survey depth over the 270 fields monitored.

In practice, scheduling constraints\footnote{Up to 20\% of the available P48 observing time is reserved for the Caltech Optical Observatory every semester.} prevented us from observing the first set of 1-d cadence fields, those that were furthest west, during the third lunation of the experiment, 2017 Oct.\ 19 -- Nov.\ 10. These fields were replaced by 90 new fields, which were observed with a 3-d cadence during the third lunation. This change in fields did not significantly affect the science output from the experiment, though it resulted in shorter duration P48 light curves for a subset of the newly discovered transients. 

\section{Observations of \fnm}

\subsection{Discovery}

\begin{figure}
\centerline{\includegraphics[width=3.5in]{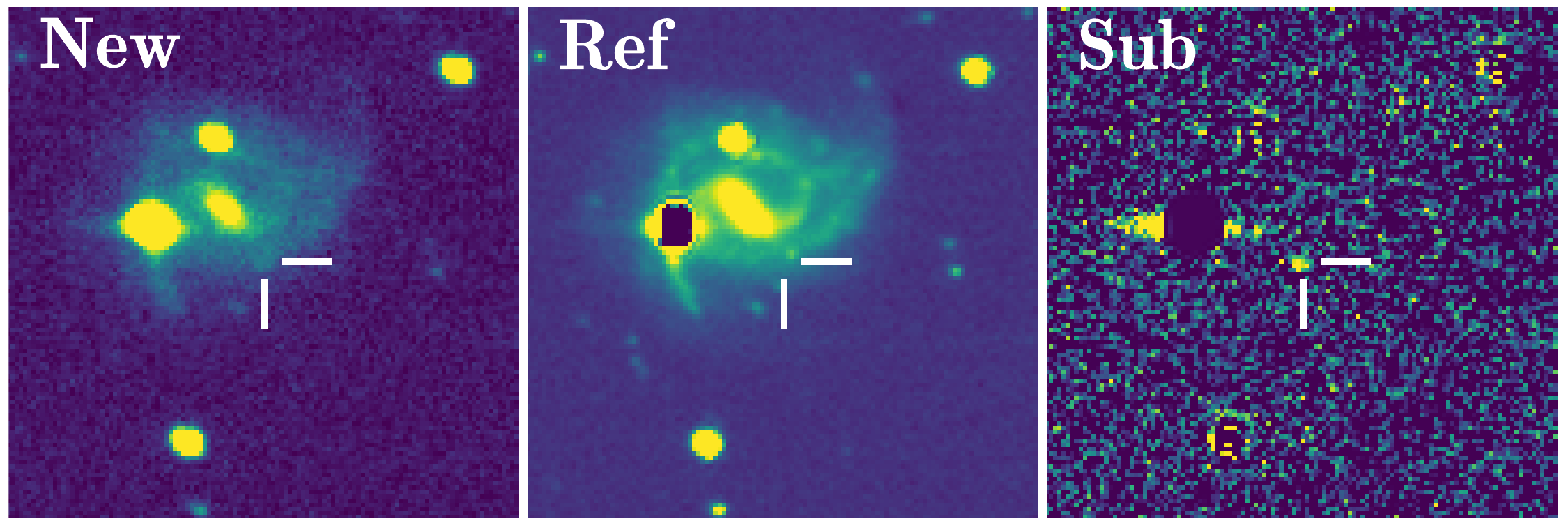}}
\caption[]{$R$-band discovery image of \fnm\ showing the new, reference, and subtraction image from left to right. Images are shown on a linear scale with the saturation levels selected to highlight faint structure in the host galaxy, UGC 00755. North is up and east to the left, and the images are centered on the SN position as indicated by the crosshairs, which are 10$\arcsec$ long. 
}
\label{fig:16fnm_finder}
\end{figure}

\fnm, located at $\alpha_{\mathrm{J2000}} = 01^{\mathrm{h}}12^{\mathrm{m}}38.32^{\mathrm{s}}$, $\delta_{\mathrm{J2000}} = +38^{\circ}30^{\mathrm{m}}08$\farcs$8$, was first detected by iPTF at $R_\mathrm{PTF} = 20.23 \pm 0.17 \; \mathrm{mag}$\footnote{Photometry for \fnm\ is reported in the $g_\mathrm{PTF}$ and $R_\mathrm{PTF}$ filters throughout, which are similar to the SDSS $g'$ and $r'$ filters, respectively (see \citealt{Ofek12} for details on PTF calibration). The correction from the $g_\mathrm{PTF}$ and $R_\mathrm{PTF}$ filters to SDSS $g'$ and $r'$ requires knowledge of the intrinsic source color (see Eqns.~1 and 2 in \citealt{Ofek12}). 02cx-like SNe do not follow a standard spectral evolution, so the color terms for \fnm\ are unknown.} on 2016 Aug.\ 26.47 (UT dates are used throughout this paper). Following automated processing \citep{Cao16,Masci17} and the use of machine learning software that separates real transients from image subtraction artefacts, known as \texttt{realbogus} \citep{Bloom12,Brink13,Masci17}, the candidate was manually saved and internally designated \fnm. The discovery image of \fnm\ is shown in Figure~\ref{fig:16fnm_finder}.

The \fnm\ host galaxy, UGC 00755, has a redshift $z_\mathrm{host} = 0.02153$ \citep{Wegner93}. Adopting $H_0 = 73~\mathrm{km~s}^{-1}~\mathrm{Mpc}^{-1}$, and correcting for Virgo infall, this redshift corresponds to a distance $d = 90.5\pm6.3~\mathrm{Mpc}$ and distance modulus $\mu = 34.78\pm0.15~\mathrm{mag}$ \citep{Mould00}. Thus, at the time of discovery \fnm\ had an absolute magnitude $M_{R_\mathrm{PTF}} \approx -14.5~\mathrm{mag}$. The light curve peaked $\sim$3 days later at $M \approx -15~\mathrm{mag}$ in both the $g_\mathrm{PTF}$ and $R_\mathrm{PTF}$ filters, suggesting \fnm\ may be a ``gap'' transient (see \citealt{Kasliwal11b,Kasliwal12a} and references therein).

The first spectrum of \fnm\ was obtained on 2016 Sep.\ 03.25 with the SEDm integral field unit (IFU) spectrograph (\citealt{Ben-Ami12}, Blagorodnova et al.\ 2017, in prep.) on the Palomar 60-inch telescope (P60). This spectrum showed low-velocity \ion{Si}{2} absorption, which in conjunction with the relatively faint absolute magnitude suggested that \fnm\ may be a sub-luminous 02cx-like SN, similar to \ha\ \citep{Foley09}. Subsequent spectra obtained with larger aperture telescopes confirmed this initial classification. 

\subsection{Photometry}

\begin{figure}
\centerline{\includegraphics[width=3.5in]{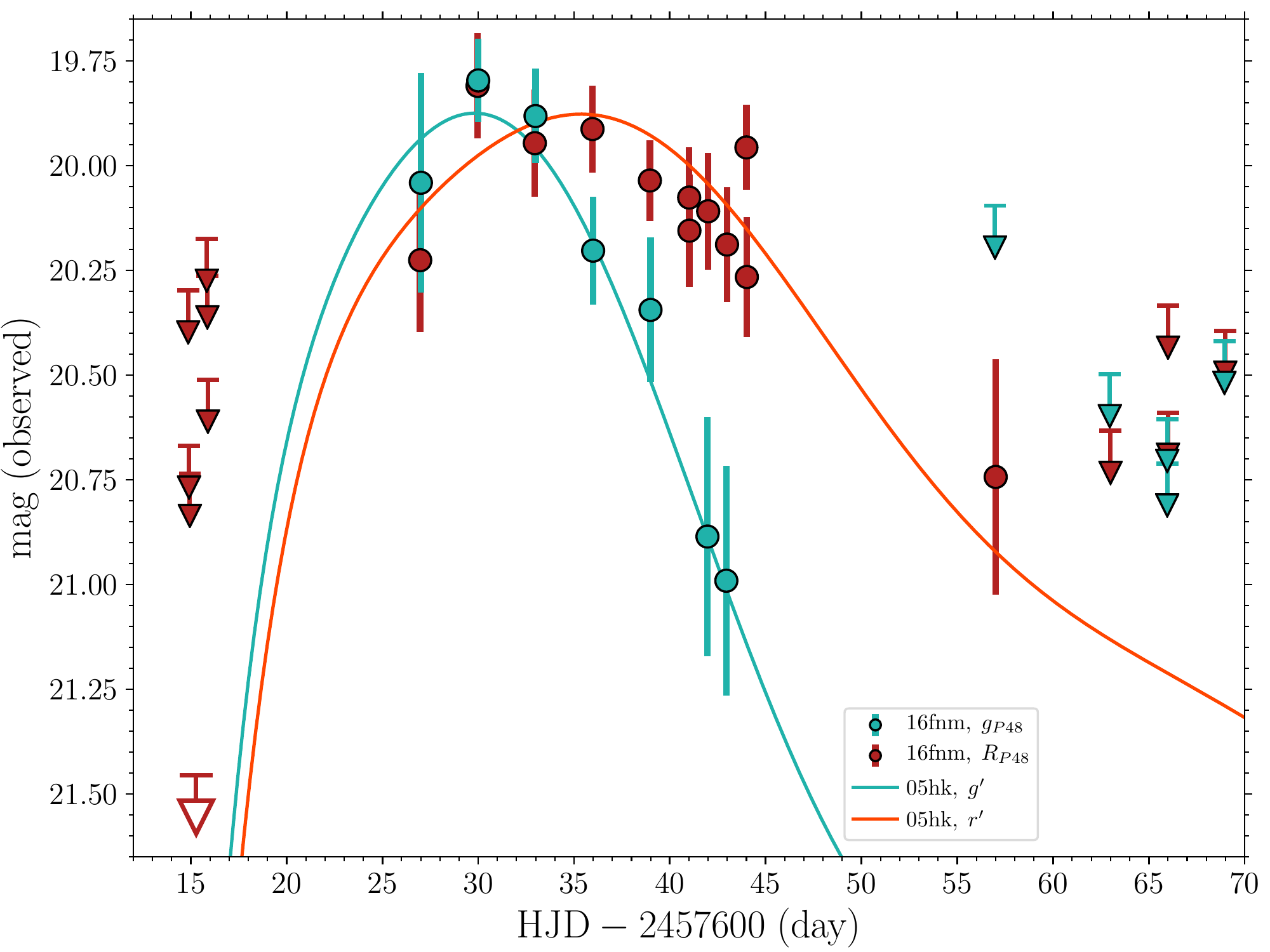}}
\caption[]{P48 light curves showing the evolution of \fnm. Teal and crimson circles show detections in the $g_\mathrm{PTF}$ and $R_\mathrm{PTF}$ bands, respectively. $5\sigma$ upper limits are shown with downward arrows. The open downward arrow shows the $5\sigma$ upper limit from combining the 6 epochs taken $\sim$11-12 d prior to discovery. The solid teal and red-orange lines show polynomial fits to the $g'$ and $r'$ filter observations of SN~2005hk \citep{Phillips07}. The curves have been shifted to align the time and brightness of maximum in $g$ band for both SNe and stretched to match the redshift of \fnm. The general agreement between the two suggests the rise time for \fnm\ is similar to SN~2005hk, $\sim$15 d.
}
\label{fig:16fnmLC}
\end{figure}

Photometric observations of \fnm\ were conducted in the $g_\mathrm{PTF}$ and $R_\mathrm{PTF}$ bands using the PTF camera \citep{law09} on the Palomar 48-inch telescope. The brightness of \fnm\ was measured following image subtraction via point-spread-function (PSF) photometry with the \texttt{PTFIDE} software package \citep{Masci17}. These measurements are summarized in Table~\ref{tbl:phot}, and shown in Figure~\ref{fig:16fnmLC}. We only consider the SN detected in epochs where the signal-to-noise ratio (SNR) is $\ge 3$, while we otherwise conservatively report $5\sigma$ upper limits.

\begin{deluxetable}{crr}
\tabletypesize{\small}
\tablewidth{2in}
\setlength{\tabcolsep}{220pt}
\tablecolumns{3}
\tablecaption{\fnm\ P48 photometry\label{tbl:phot}}
\tablehead{ \colhead{$\mathrm{HJD} - 2{,}457{,}600$} & \colhead{mag\tablenotemark{a}} & \colhead{$\sigma_\mathrm{mag}$}}
\startdata
\multicolumn{3}{c}{$g_\mathrm{PTF}$}\\
\hline
27.008 & 20.04 & 0.26 \\
30.002 & 19.80 & 0.10 \\
32.990 & 19.88 & 0.11 \\
35.998 & 20.20 & 0.13 \\
38.992 & 20.34 & 0.17 \\
41.964 & 20.89 & 0.29 \\
42.951 & 20.99 & 0.27 \\
56.972 & $>$20.10 & ... \\
62.963 & $>$20.50 & ... \\
65.962 & $>$20.71 & ... \\
65.963 & $>$20.60 & ... \\
68.951 & $>$20.42 & ... \\
\hline
\multicolumn{3}{c}{$R_\mathrm{PTF}$}\\
\hline
14.863 & $>$20.30 & ... \\
14.905 & $>$20.67 & ... \\
14.948 & $>$20.74 & ... \\
15.833 & $>$20.17 & ... \\
15.863 & $>$20.26 & ... \\
15.892 & $>$20.51 & ... \\
26.971 & 20.23 & 0.17 \\
29.964 & 19.81 & 0.13 \\
32.954 & 19.95 & 0.13 \\
35.964 & 19.91 & 0.10 \\
38.960 & 20.04 & 0.10 \\
41.009 & 20.08 & 0.12 \\
41.022 & 20.16 & 0.13 \\
42.002 & 20.11 & 0.14 \\
42.986 & 20.19 & 0.14 \\
43.998 & 19.96 & 0.10 \\
44.019 & 20.27 & 0.14 \\
57.008 & 20.74 & 0.28 \\
63.001 & $>$20.63 & ... \\
66.001 & $>$20.59 & ... \\
66.002 & $>$20.33 & ... \\
68.991 & $>$20.39 & ...
\enddata
\tablenotetext{a}{Observed mag, not corrected for Galactic extinction. $5\sigma$ upper limits are reported for epochs where \fnm\ is not detected.}
\end{deluxetable}

We determine the time of and brightness at maximum for \fnm\ via second-order polynomial fits to the epochs when \fnm\ was detected. From these fits we find that \fnm\ peaked on HJD~$2{,}457{,}629.8 \pm 2.0$ and $2{,}457{,}632.8 \pm 4.8$ at $19.87 \pm 0.07~\mathrm{mag}$ and $19.88 \pm 0.05~\mathrm{mag}$, in the $g_\mathrm{PTF}$- and $R_\mathrm{PTF}$-bands, respectively. These measurements represent the observed maxima of \fnm, and have not been corrected for Milky Way or host galaxy extinction. 02cx-like SNe do not follow a standard color evolution, meaning host-galaxy extinction cannot be inferred from photometry alone. We do not detect narrow \ion{Na}{1}~D at the redshift of UGC 00755 in any of our spectra, and therefore we assume that host-galaxy extinction is negligible. This assumption is supported by the observed blue color of \fnm\ at peak. 

Adopting the distance modulus to UGC 00755, and correcting for foreground Galactic extinction ($A_g = 0.178$ mag, $A_r = 0.123$ mag; \citealt{Schlafly11}), we find that \fnm\ peaked at $M_{g_\mathrm{PTF}} = -15.09 \pm 0.17$ mag and $M_{R_\mathrm{PTF}} = -15.02 \pm 0.16$ mag, under the assumption of no local host extinction.

We cannot place strong observational constraints on the rise time of \fnm: prior to its initial detection, iPTF had not observed this field for $\sim$12 d (see Figure~\ref{fig:16fnmLC}). If we combine the forced-PSF flux measurements from the 6 epochs taken between HJD~$2{,}457{,}614$ and $2{,}457{,}616$, we derive a more constraining inverse-variance-weighted upper limit of $R_\mathrm{PTF} > 21.46~\mathrm{mag}$ on HJD~$2{,}457{,}615.3$. This deeper non-detection suggests the rest-frame rise time of \fnm\ is $< 17.2~\mathrm{d}$ in the $R$-band. Further constraints on the rise-time are available by comparing \fnm\ and SN~2005hk, the 02cx-like SN with the best observational constraints on the time of explosion \citep{Phillips07}. Figure~\ref{fig:16fnmLC} shows polynomial fits to the $g'$- and $r'$-band light curves of SN~2005hk, shifted to match the time and brightness of \fnm\ in the $g_\mathrm{PTF}$ band and stretched to match the redshift of \fnm. Formally, no further stretch factor is required to provide an excellent match between SN~2005hk and \fnm, as can be seen in Figure~\ref{fig:16fnmLC}. Assuming that SN~2005hk and \fnm\ have similar compositions, opacities, and temperatures, which is reasonable based on the spectral similatities of the two supernovae, then the rise time of \fnm\ is $\sim{15}~\mathrm{d}$ \citep{Phillips07}, consistent with our $R_\mathrm{PTF}$ observational constraints above.

Furthermore, a gap in our observations between $\sim$13 and 27 d after $g$-band maximum make it difficult to properly measure the decline rate of \fnm. If we adopt the same quadratic fit used to determine the time of $g$-band maximum to estimate the SN brightness beyond $+13~\mathrm{d}$ after maximum, then \fnm\ declined by $1.75 \pm 0.46~\mathrm{mag}$ in the $g_\mathrm{PTF}$ band after $15~\mathrm{d}$ in its rest frame. Using the same similarity arguments about SN~2005hk from above, then we would expect the two SNe to feature similar declines, meaning for \fnm\ $\Delta m_{15}(B) \approx 1.6~\mathrm{mag}$, as was found for SN~2005hk \citep{Phillips07}. 

\subsection{Spectroscopy}

\begin{figure*}
\centerline{\includegraphics[width=6.5in]{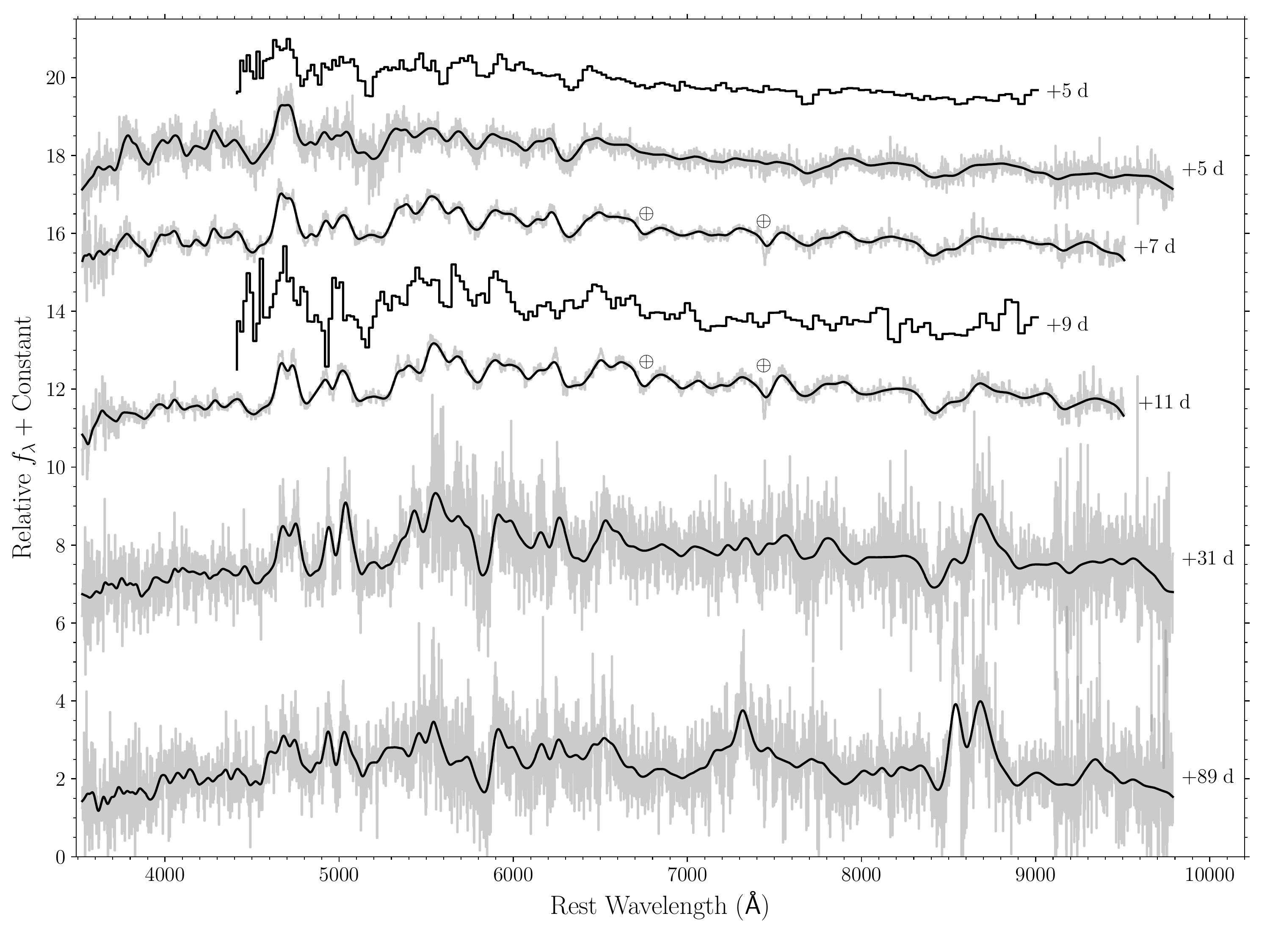}}
\caption[]{Spectral evolution of \fnm. The spectra are labeled with their ages in the rest frame of the SN relative to the observed $g_\mathrm{PTF}$-band maximum on 2016 August 29.3. The raw spectra are shown in grey, while the solid black lines are convolved with a Gaussian filter with $\mathrm{FWHM} = 2{,}500 \; \mathrm{km \; s}^{-1}$. The low SNR portion of SEDm spectra ($+5$ and $+9$) is trimmed blueward of 4500 \AA\ in the observer frame, and these spectra are not smoothed. From the first spectrum, \fnm\ exhibited low-velocity features characteristic of 02cx-like SNe.
}
\label{fig:16fnmSpec}
\end{figure*}

Optical spectra of \fnm\ were obtained on 2016 September 03.3 and 2016 September 07.3 with the SEDm on the Palomar 60-in telescope (P60). Additional spectra were obtained on 2016 September 03.4 with the double beam spectrograph (DBSP; \citealt{oke95}) on the Palomar 200-in telescope (P200), on 2016 September 06.1 and September 10.1 with the Andalucia Faint Object Spectrograph and Camera (ALFOSC) on the Nordic Optical Telescope (NOT), and on 2016 September 30.5 and November 28.4 with the low-resolution imaging spectrometer (LRIS; \citealt{Oke82}) on the Keck I 10 m telescope. All spectra were extracted and calibrated using standard procedures. The sequence of \fnm\ spectra is shown in Figure~\ref{fig:16fnmSpec}. A log of our spectroscopic observations is presented in Table~\ref{tbl:spec}.\footnote{Our \fnm\ spectra will be publically released via WISeREP \citep{Yaron12} following paper acceptance.}

\begin{deluxetable}{ccclrc}
\tabletypesize{\small}
\setlength{\tabcolsep}{160pt}
\tablecolumns{6}
\tablecaption{Log of Spectroscopic Observations\label{tbl:spec}}
\tablehead{ \colhead{t\tablenotemark{a}} & 
            \colhead{UT Date} & 
            \colhead{Instrument\tablenotemark{b}} & 
            \colhead{Range} & 
            \colhead{Exp\tablenotemark{c}} & 
            \colhead{Air}  \\
            \colhead{(d)} &
            \colhead{} & 
            \colhead{} & 
            \colhead{(\AA)} & 
            \colhead{(s)} & 
            \colhead{Mass}
            }
\startdata
~4.9 & 2016-09-03.25 & SEDm & 3806--9187 & 2700  & 1.51  \\
~5.0 & 2016-09-03.36 & DBSP & 3101--10236 & 1200  & 1.07  \\
~7.7 & 2016-09-06.13 & ALFOSC & 3427--9714 & 4800  & 1.02  \\
~8.8 & 2016-09-07.25 & SEDm & 3807--9187 & 2700  & 1.46  \\
11.5 & 2016-09-10.06 & ALFOSC & 3556--9712 & 4800  & 1.10  \\
31.5 & 2016-09-30.53 & LRIS & 3071--10297 & 975 & 1.14  \\
89.1 & 2016-11-28.37 & LRIS & 3057--10276 & 3370 & 1.14
\enddata
\tablenotetext{a}{Age in rest-frame days relative to the observed $g_\mathrm{PTF}$-maximum on 2016-08-29.298.}
\tablenotetext{b}{SEDm: Spectral Energy Distribution Machine on the Palomar 60-in telescope. DBSP: Double Beam Spectrograph on the 200-in Palomar Hale Telescope. ALFOSC: Andalucia Faint Object Spectrograph and Camera on the 2.6-m Nordic Optical Telescope. LRIS: low-resolution imaging spectrograph on the 10-m Keck-I telescope.}
\tablenotetext{c}{Exposure time.}
\end{deluxetable}

\fnm\ shows the hallmark features of 02cx-like SNe: low velocity lines of intermediate-mass and Fe-group elements. In our earliest spectroscopic observations, taken within a week of $g_\mathrm{PTF}$-band maximum, \ion{Si}{2} $\lambda6335$ shows an expansion velocity of only $\sim$2,000 \kms. Other relatively unblended lines, such as Ca H\&K, \ion{Fe}{2} $\lambda4555$ and \ion{O}{1} $\lambda7774$ show expansion speeds of $\sim$3,200 \kms. 

The late time spectra of \fnm\ resemble the evolution of SN~2008ha. While the [\ion{Ca}{2}] $\lambda\lambda$7291, 7323 and \ion{Ca}{2} NIR triplet are not as strong in \fnm\ as in SN~2008ha, they are clearly present, unlike in SN~2002cx. Furthermore, like SN~2008ha, absorption from \ion{Fe}{2} $\lambda$4555 is prominent more than a month after maximum. Finally, both \fnm\ and SN~2008ha exhibit a strong P Cygni profile at $\sim$5800 \AA, possibly associated Na~D, for months after maximum light, even as the rest of the optical spectrum transitions to a nebular state.

\begin{figure}
\centerline{\includegraphics[width=3.5in]{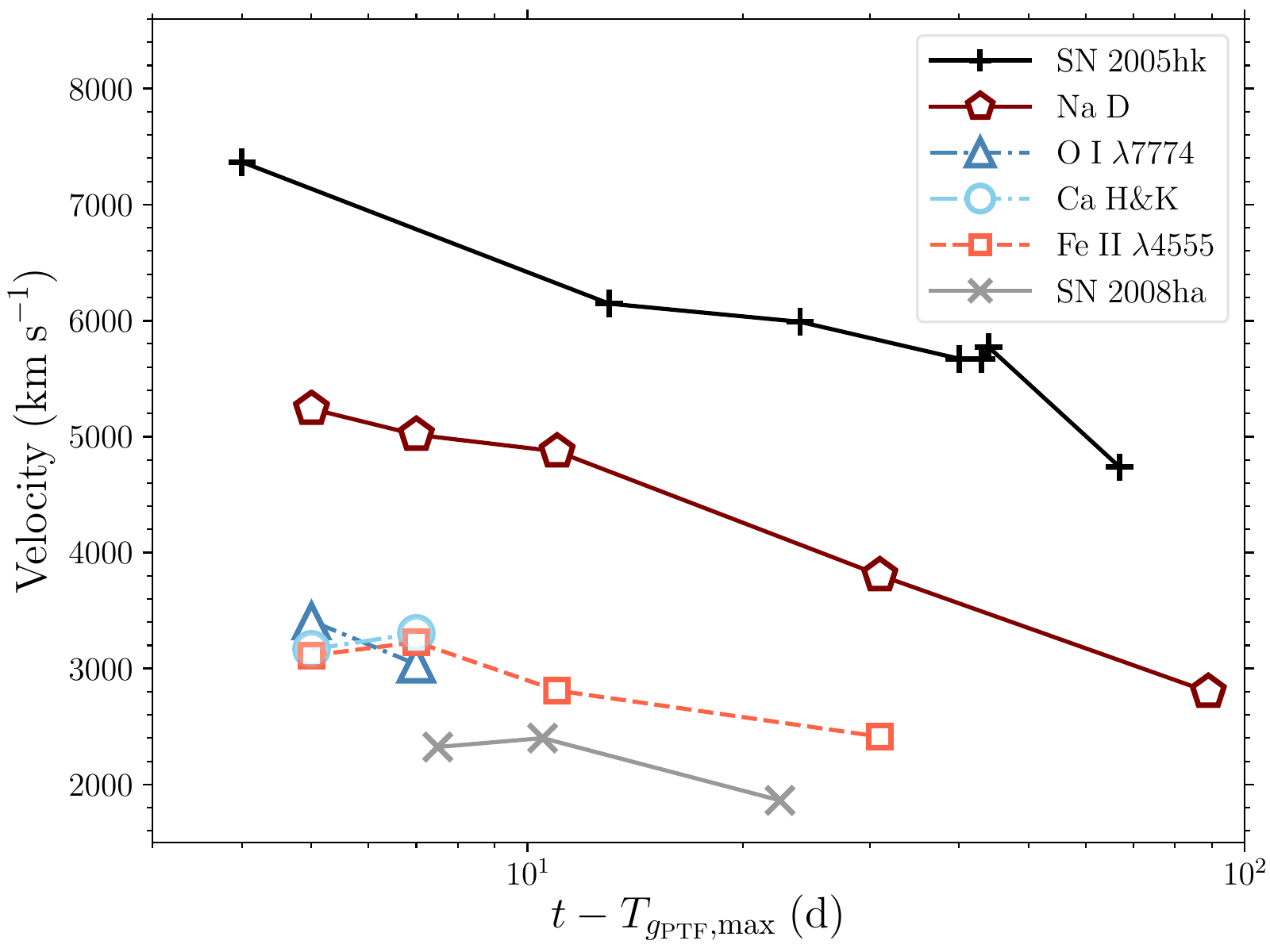}}
\caption[]{Velocity evolution of the absorption minimum of Ca H\&K (circles), \ion{Fe}{2} $\lambda{4445}$ (squares), Na D (pentagons), \ion{O}{1} $\lambda{7774}$ (triangles) for \fnm\ traced by the minimum of absorption. Also shown is the velocity evolution of SN~2005hk (pluses, as traced by \ion{Fe}{2} $\lambda{4445}$) and SN~2008ha (X symbols, as traced by \ion{O}{1} $\lambda{7774}$). The spectra for SN~2005hk and SN~2008ha are taken from \citet{Phillips07} and \citet{Foley09}, respectively. \fnm\ is clearly intermediate between SN~2005hk and SN~2008ha. For \fnm\ the Na D doublet may be contaminated by other features affecting our velocity measurements.
}
\label{fig:16fnmVelEvolution}
\end{figure}

We measure the expansion velocities for different species in \fnm\ by tracking the wavelength of the minimum for each feature, as shown in Figure~\ref{fig:16fnmVelEvolution}. The measurements are made following a convolution of the spectra with a gaussian kernel with full-width half-max (FWHM) $= 2{,}000$ \kms. Blending and the relatively low SNR of the spectra make it challenging to track the evolution of individual lines more than 10 days after $g_\mathrm{PTF}$-band maximum. The \ion{Fe}{2} $\lambda$4555 feature shows a modest decrease of only $\sim$700 \kms\ from $+5$ to $+31$ d. At $+5$ d after maximum the Na~D feature exhibits significantly faster speeds of $\sim$5,200 \kms, decreasing to $\sim$2,800 \kms\ at $+89$ d. The higher velocities for this line indicates that our measurement may be contaminated by other features. Taken together, these lines demonstrate that \fnm\ has a velocity structure that is intermediate between SN~2005hk, with typical velocities of $\sim$7,000 \kms\ \citep{Phillips07}, and SN~2008ha, with typical velocities of $\sim$2,000 \kms\ \citep{Foley09, Valenti09}.

\section{Comparison to other sub-luminous 02cx-like SNe}

\begin{figure}
\centerline{\includegraphics[width=3.5in]{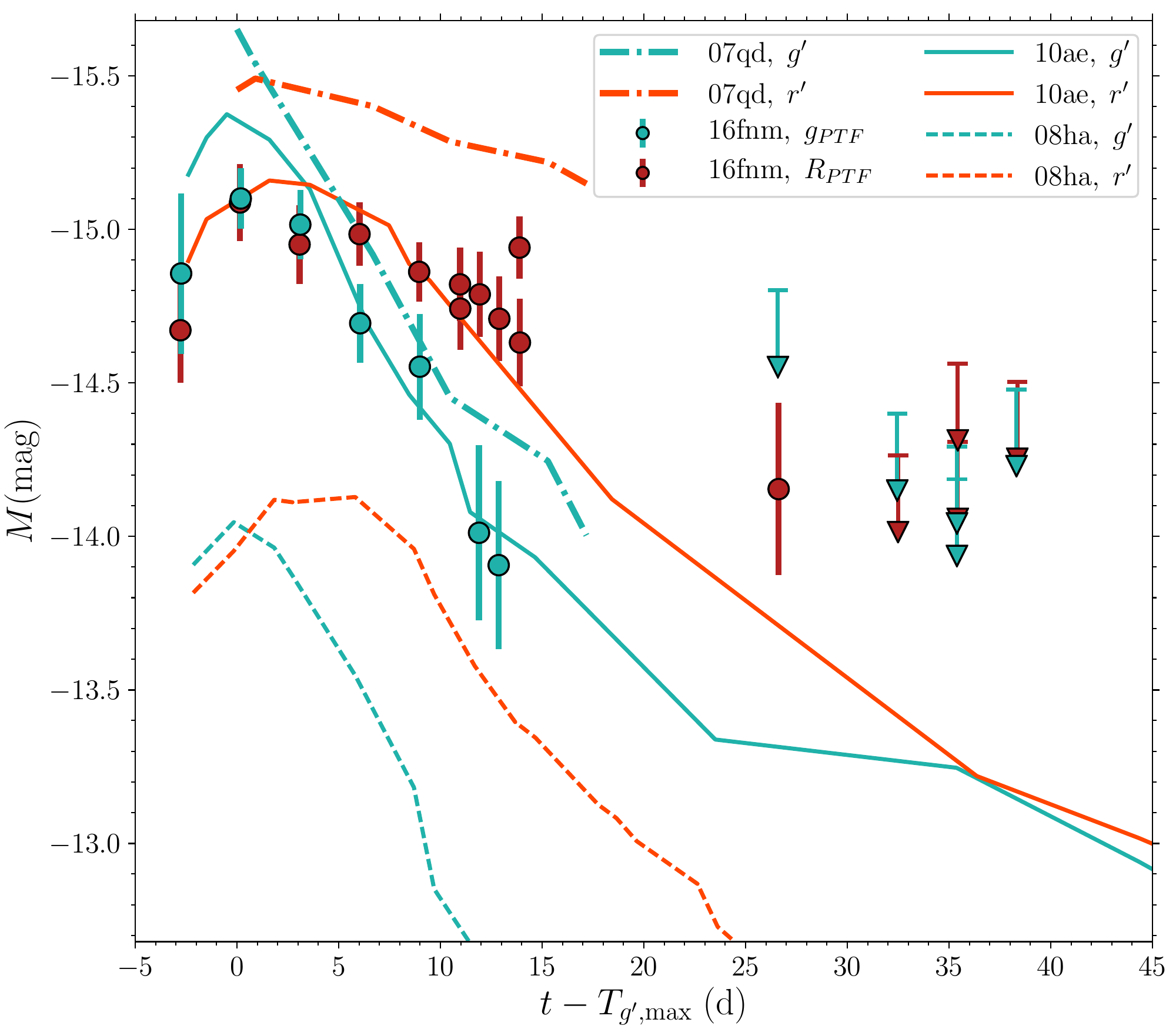}}           
\caption[]{Photometric evolution of the faintest members of the 02cx-like class in the $g'$ and $r'$ filters. From brightest to faintest they are: SN~2007qd, SN~2010ae, \fnm, and SN~2008ha. These SNe all feature fast declines in the $g'$ band, $\Delta m_{15}(g) \ga 1.2 \; \mathrm{mag}$. SN~2007qd and \fnm, on the other hand, exhibit relatively slow declines in the $r'$ filter.
}
\label{fig:LCcomp}
\end{figure}

We compare the photometric evolution of \fnm\ to other sub-luminous 02cx-like SNe in Figure~\ref{fig:LCcomp}. The comparison SNe are: SN~2007qd \citep{McClelland10}, SN~2010ae \citep{Stritzinger14}, and SN~2008ha \citep{Stritzinger14}, all low luminosity 02cx-like SNe with $g'$- and $r'$-band photometric coverage.\footnote{We remind the reader that \fnm\ was observed in the $g_\mathrm{PTF}$ and $R_\mathrm{PTF}$ filters, which are similar to SDSS $g'$ and $r'$ (see \citealt{Ofek12} for the filter transformations).} Each light curve in Figure~\ref{fig:LCcomp} has been corrected for foreground Galactic extinction using the \citet{Schlafly11} updates to the \citet{Schlegel98} reddening maps. SN~2010ae has been further corrected for a host galaxy reddening of $E(B - V)_\mathrm{host} = 0.50 \, \mathrm{mag}$ \citep{Stritzinger14,Foley13}. The light curves have been shifted to align the time of $g'$-band maximum, where we have assumed the first detection of SN~2007qd corresponds to $g'$-band maximum (see \citealt{McClelland10} for further details). We adopt distance moduli of 36.23, 30.44, and 31.64 mag for SN~2007qd, SN~2010ae, and SN~2008ha, respectively, which are corrected for Virgo infall \citep{Mould00} and assume $H_0 = 73 \, \mathrm{km} \, \mathrm{s}^{-1} \, \mathrm{Mpc}^{-1}$. No $K$- or $S$-corrections have been applied.

Figure~\ref{fig:LCcomp} shows that \fnm\ is generally similar to SN~2010ae and SN~2007qd. In particular, all 3 SNe are of comparable brightness at peak ($15 \, \mathrm{mag} \lesssim M \lesssim 15.5 \, \mathrm{mag}$), and exhibit a fast decline in the $g'$ band. SN~2008ha, on the other hand, is significantly fainter than the other SNe. Interestingly, both \fnm\ and SN~2007qd exhibit a slow decline in the $R_\mathrm{PTF}/r'$ band, evolving from $g' - r' \approx 0.1 \, \mathrm{mag}$ near peak to $g' - r' \approx 0.9 \, \mathrm{mag}$ at $t \approx 15 \, \mathrm{d}$.

\begin{figure*}
\centerline{\includegraphics[width=6in]{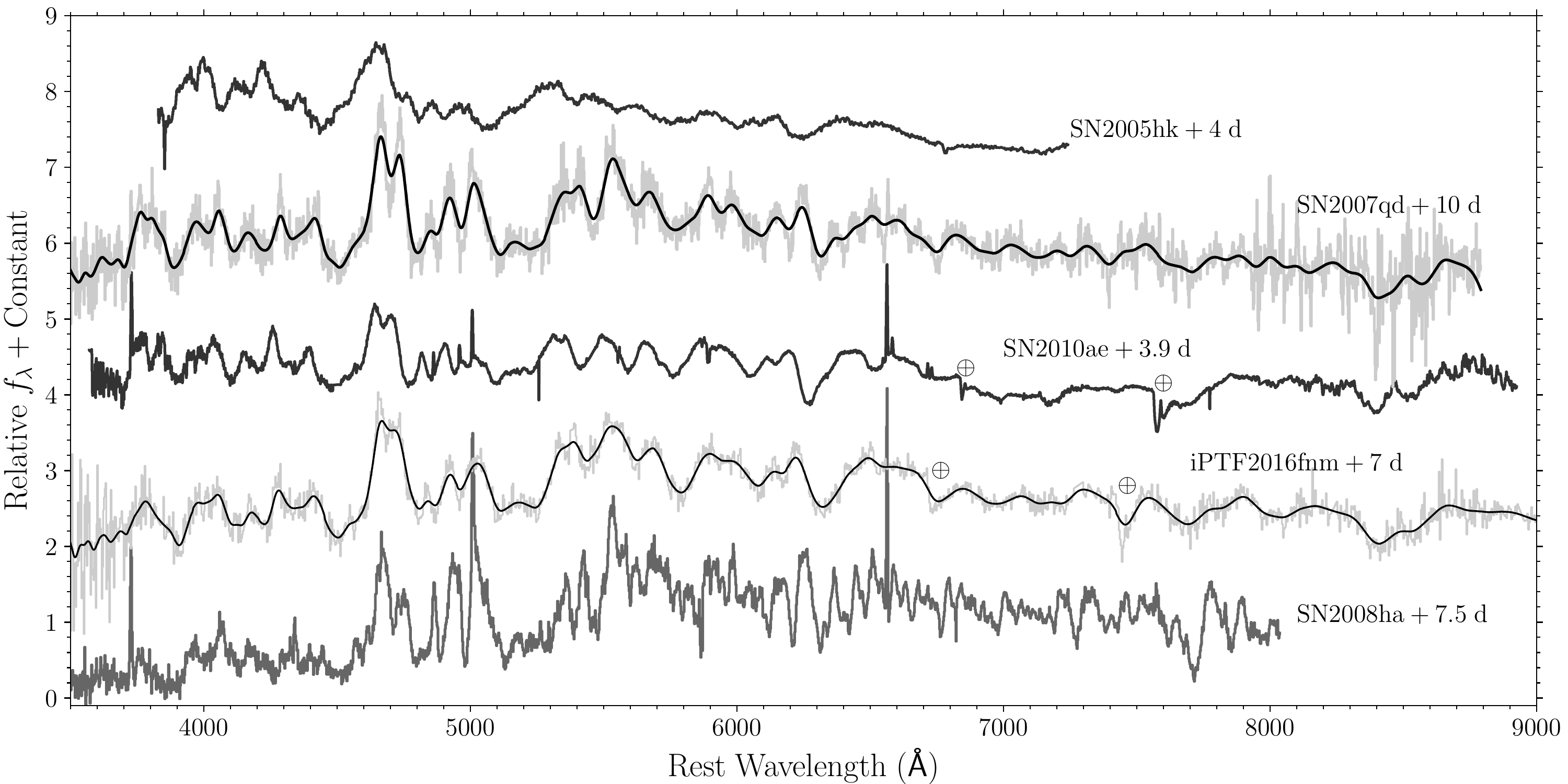}}
\caption[]{Spectral comparisons of \fnm\ to other subluminous 02cx-like SNe. The spectra are labeled with their ages in the rest frame of the SN relative to $B$- or $g'$-band maximum. SN~2007qd and \fnm\ are also shown following convolution with a $\mathrm{FWHM} = 2,500 \; \mathrm{km \; s}^{-1}$ gaussian kernel. From top to bottom, the spectra are SN~2005hk \citep{Phillips07}, SN~2007qd \citep{McClelland10}, SN~2010ae \citep{Stritzinger14}, \fnm, and SN~2008ha \citep{Foley09}. SN~2005hk is a prototypical 02cx-like SN, while SNe~2007qd and 2010ae, like \fnm, were fainter at the time of maximum brightness with lower expansion velocities. SN~2008ha is the faintest member of the 02cx-like class.
}
\label{fig:16fnmSpec_comp}
\end{figure*}

We compare the same 4 SNe, as well as SN~2005hk, at similar epochs, $t \approx +4$ to $+10$ d, in Figure~\ref{fig:16fnmSpec_comp}. The spectra are ordered top to bottom from the most luminous to least luminous, though note that changes in host-galaxy reddening could re-arrange the middle 3 spectra. From Figure~\ref{fig:16fnmSpec_comp} it is clear that SN~2005hk has the highest velocity features, while SN~2007qd, SN~2010ae, and \fnm\ all have similar velocities, and SN~2008ha has the lowest velocity signatures.\footnote{While these 5 SNe show an apparent correlation between absolute magnitude and velocity for 02cx-like SNe, SN~2009ku, the only 02cx-like SN with velocities as low as SN~2008ha, is also the most luminous 02cx-like SN \citep{Narayan11}.} The overall similarity of SN~2007qd, SN~2010ae, and \fnm\ is striking. These 3 SNe are clearly closely related with similar velcities and ejecta composition, which is dominated by intermediate mass elements. While the 02cx-class as a whole exhibits great diversity, SN~2007qd, SN~2010ae, and \fnm\ feature nearly identical spectra and photometric evolution suggesting they had similar progenitors or explosion mechanisms.

\section{Results}\label{sec:results}

While 02cx-like SNe exhibit a large range in luminosity (roughly 2 orders of magnitude separate SN~2008ha and SN~2009kr), their general similarity to normal Type Ia SNe have led many to search for correlated observational properties to unify the class. While the precise details of the explosion mechanism are debated, the observational consensus points to white dwarf (WD) progentiors for 02cx-like SNe (see e.g., \citealt{Foley13,Stritzinger15,Magee17}). Thus, it may be the case that 02cx-like SNe are a single parameter family, much like type Ia SNe, whose evolution is largely controlled by the amount of $^{56}$Ni synthesized during explosion \citep{Mazzali07}.

Using the first four known 02cx-like SNe, SN~2002cx, SN~2005hk, SN~2007qd, and SN~2008ha, a potential correlation between absolute magnitude and ejecta velocity for 02cx-like SNe, whereby more luminous events also have faster ejecta, was identified in \citet{McClelland10}. The subsequent discovery of SN~2009ku \citep{Narayan11}, which had extremely low velocities, like SN~2008ha, while also being the most luminous member of the class, poses a significant challenge to this correlation. While an increased sample of 02cx-like SNe shows a general correlation between ejecta velocity and luminosity (see Figure~20 in \citealt{Foley13}), SN~2009ku remains a significant outlier. 

\subsection{The Luminosity-Decline Relation for 02cx-like SNe}

Many studies have examined whether 02cx-like SNe follow their own Phillips relation, similar to type Ia SNe, whereby more luminous SNe Ia also have broader light curve shapes \citep{Phillips93}. Using a sample of 13 02cx-like SNe, evidence for an anti-correlation between $M_V$ and $\Delta{m_{15}}(V)$ is presented in \citet{Foley13}. The $V$-band $M-\Delta{m}_{15}$ relation presented in \citet{Foley13} exhibited significant scatter, well in excess of that found for normal type Ia SNe. Furthermore, the sample in \citet{Foley13} excludes SN~2007qd, which declined at a similar rate as SN~2005hk, despite being $\sim$2.5 mag fainter at peak.\footnote{SN~2007qd was first detected after $g'$-band maximum, which is why it was excluded from the sample in \citet{Foley13}. However, the 50 d range for the time of maximum for SN~2007qd in Table~5 of \citet{Foley13} ignores the deep upper limits reported in \citet{McClelland10} between $\sim$10--20 d prior to SN~2007qd's first detection. As argued in \citet{McClelland10}, the initial detection of SN~2007qd is very likely near the epoch of $g'$ maximum.}

\begin{figure}
\centerline{\includegraphics[width=3.5in]{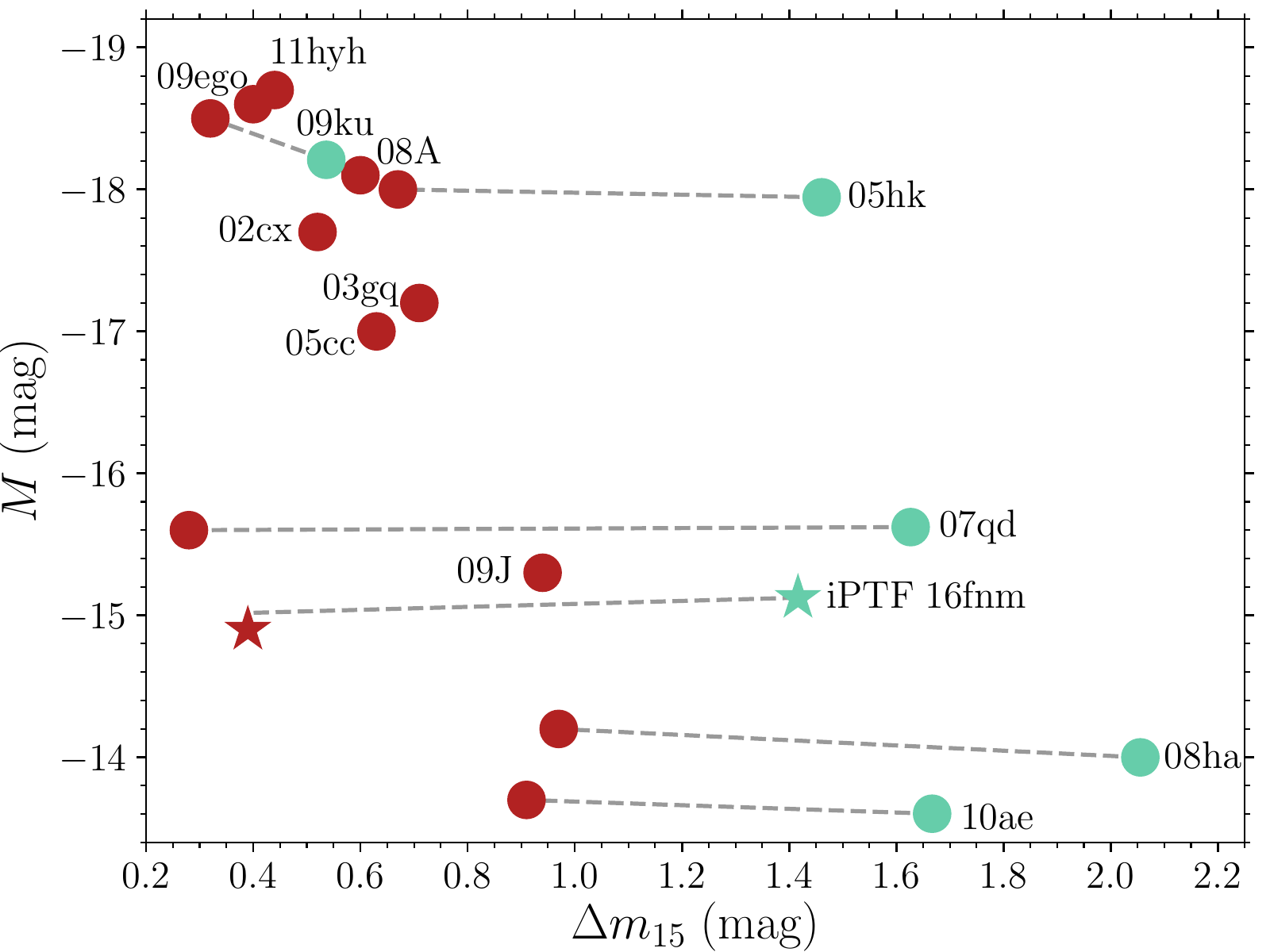}}
\caption[]{Absolute magnitude vs.\ $\Delta m_{15}$, both in the $r'$ (shown in crimson) and $g'$-bands (teal) for 02cx-like SNe, adapted from \citet{White15}. The dashed grey lines connect SNe for which both $r'$ and $g'$ observations are available. Note that following \citet{White15}, we have not corrected $M$ for host galaxy extinction, which is why SN~2010ae is fainter than SN~2008ha in this figure. In the $r'$ band most 02cx-like SNe lie along a high-scatter sequence, though SN~2007qd and \fnm\ are clear outliers. Similarly, $g'$-band observations of 02cx-like SNe also show weak evidence for a simple $M-\Delta{m}_{15}$ relation. 
}
\label{fig:16fnm_deltaM15}
\end{figure}

\fnm\ provides additional evidence for the lack of a simple $M-\Delta{m}_{15}$ relation for 02cx-like SNe. In particular, Figure~\ref{fig:16fnmLC} shows that \fnm\ declined at a nearly identical rate as SN~2005hk, which peaked at $M_{g'} = -18.08 \pm 0.25 \; \mathrm{mag}$ \citep{Stritzinger15}, 3 mag brighter than \fnm, $M_{g_\mathrm{PTF}} = -15.08 \pm 0.17 \; \mathrm{mag}$. To further illustrate this point, we update Figure~10 from \citet{White15} to include \fnm, as shown in Figure~\ref{fig:16fnm_deltaM15}. For 02cx-like SNe with available $g'$ observations, we additionally include estimates of $M_{g'}$ and $\Delta{m_{15}}(g')$ using the same procedure adopted in \citet{White15}. Even if SN~2007qd is excluded from the sample, \fnm\ stands out as a strong outlier for any $M-\Delta{m}_{15}$ relation in the $r'/R_{PTF}$-band. Examining those sources with $g'$ observations, Figure~\ref{fig:16fnm_deltaM15} shows weak evidence for a $M-\Delta{m}_{15}$ relation, though the scatter is extremely large. As already noted, SN~2005hk and \fnm\ exhibit similar decline rates but differ by $\sim$3 mag at peak. Furthermore, SN~2005hk and SN~2009ku both peak at $M_{g'}\approx{-18} \; \mathrm{mag}$, yet their $\Delta{m_{15}}(g')$ differ by $\sim$1 mag. 

While a larger sample with better photometric coverage is still required, we conclude that the 02cx-like class of SNe cannot be well described by a single $M-\Delta{m}_{15}$ relation. As future time-domain surveys significantly increase the sample of known 02cx-like SNe, it may be the case that further subdivision of the 02cx-like class yields a subset that can be described by a simple $M-\Delta{m}_{15}$ relation. 

\subsection{A Selection Function for 02cx-like SNe: $g' - r'$ Color Evolution}

In \citet{Foley13}, the color evolution of 02cx-like SNe is examined to see if all 02cx-like SNe can be described by a single color curve following a reddening correction, similar to the Lira-law for normal type Ia SNe \citep{Lira96,Phillips99}. \citeauthor{Foley13} apply reddening corrections to a sample of six 02cx-like SNe, and find a signficant reduction in scatter for the $V - R$ and $V - I$ color curves. The same corrections do not reduce the $B - V$ scatter, however. As a result, they cannot conclude whether the observed scatter is the result of dust reddening or intrinsic differences in the class.

\begin{figure*}
\centerline{\includegraphics[width=7.0in]{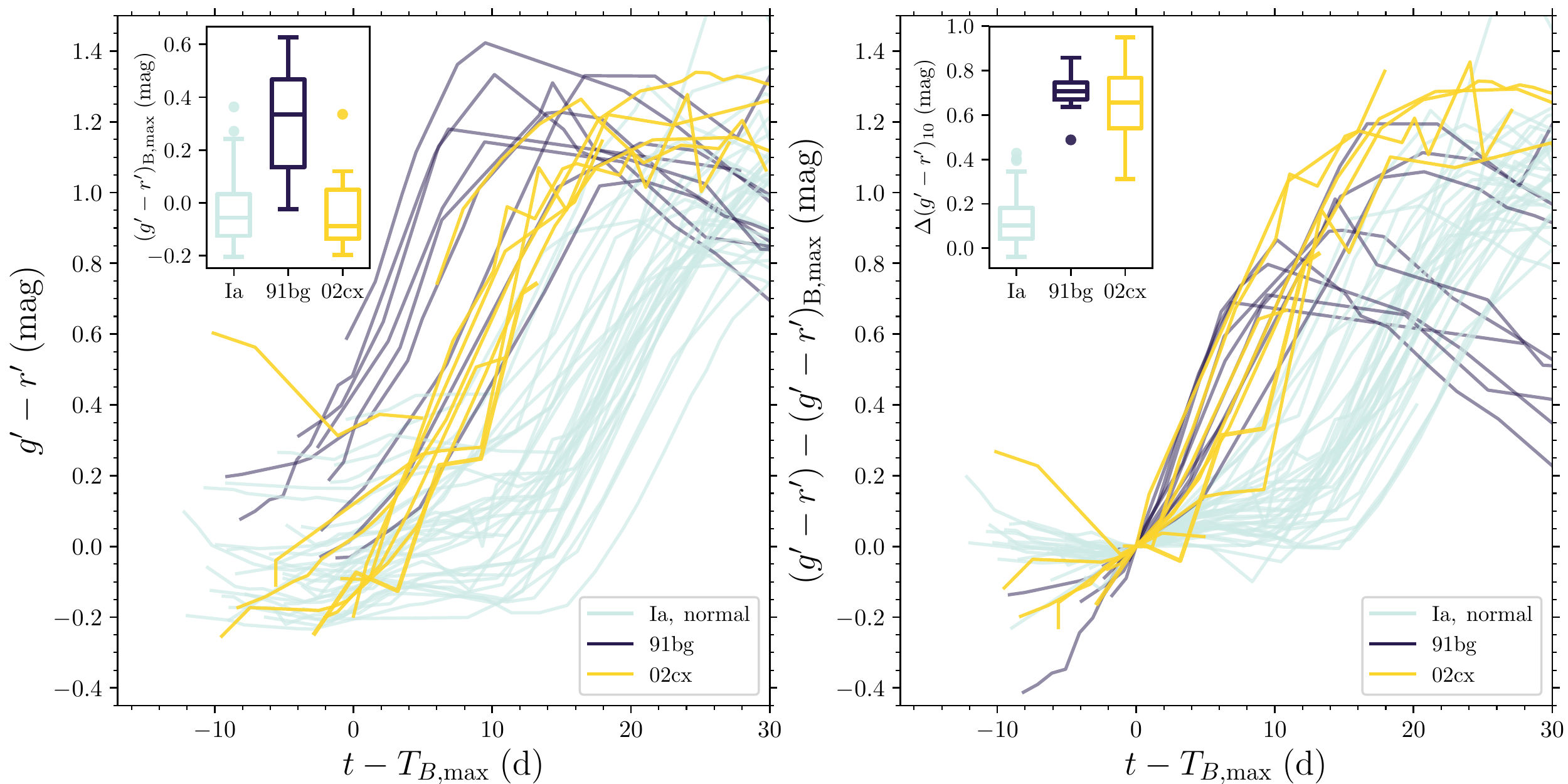}}
\caption[]{Color evolution of normal (shown in light blue), 91bg-like (purple) and 02cx-like (yellow) type Ia SNe. For clarity, photometric uncertainties are not shown and the individual measurements are connected via solid lines. The color curves are normalized to $T_{B,\mathrm{max}}$, where available (see text). The color curves are corrected for Galactic extinction, and in some cases host-galaxy extinction as well (see text). The samples for each subclass are defined in the text. The 02cx-like SNe~2009ku and 2010ae feature gaps in their $g' - r'$ color curves from $\sim$+5 d to +30 d and $\sim$+11 d to +45 d, respectively \citep{Narayan11,Stritzinger14}. As such, we only display these color curves through +5 d and +11 d, respectively.

\textit{Left}: The $g' - r'$ color evolution of type Ia SNe. The 02cx-like SNe form a remarkablely tight sequence with blue colors ($g' - r' \lesssim 0 \; \mathrm{mag}$) at peak, and a rapid decline relative to normal type Ia SNe. SN~2009ku, which has $g' - r' \approx 0.6 \; \mathrm{mag}$ at $-10 \; \mathrm{d}$, stands out as an outlier relative to the other 02cx-like SNe. The inset shows an inter-quartile range (IQR) box plot comparing the $g' - r'$ color at $T_{B,\mathrm{max}}$, determined via linear interpolation between the two epochs spanning $T_{B,\mathrm{max}}$, for the 3 subclasses. SN~2015H is excluded as the $g' - r'$ color is not available prior to +6 d \citep{Magee16}. At peak, 02cx-like SNe are very blue (the red tail of the distribution is dominated by SN~2009ku) like normal Ia SNe, while 91bg-like SNe are red.

\textit{Right}: The color evolution relative to the color at $T_{B,\mathrm{max}}$, denoted here as $(g' - r') - (g' - r')_{\mathrm{B,max}}$. By normalizing relative to the color at $T_{B,\mathrm{max}}$, we track the color evolution independent of line-of-sight extinction. The at $B$-maximum is determined via linear interpolation (see above). Both 91bg-like and 02cx-like SNe rapidly evolve to the red along a relatively tight sequence, independent of the temperature at peak. The inset IQR box plot shows the change in $(g' - r')$ color between $T_{B,\mathrm{max}}$ and +10 d, $\Delta(g' - r')_{10}$. The box plot clearly confirms that normal type Ia SNe remain blue longer than 91bg-like and 02cx-like SNe. SN~2009ku is excluded from the boxplot as there is no $(g' - r')$ measurement at +10 d \citep{Narayan11}.
}
\label{fig:gR_evolution}
\end{figure*}

Here, we instead examine the color curves of 02cx-like SNe as a possible selection function to separate them from normal type Ia SNe. While 02cx-like SNe do not follow a $M-\Delta{m}_{15}$ relation, we find that the $g' - r'$ color evolution is relatively uniform for the class.  In Figure~\ref{fig:gR_evolution}, we show the $g' - r'$ color evolution for nine 02cx-like SNe, including: SN~2008ha \citep{Stritzinger14}, SN~2010ae \citep{Stritzinger14}, SN~2007qd \citep{McClelland10}, SN~2005hk \citep{Phillips07}, SN~2009ku \citep{Narayan11}, PS1-12bwh \citep{Magee17}, SN~2015H \citep{Magee16}, SN~2012Z \citep{Stritzinger15}, and \fnm\ (where we are using $g_\mathrm{PTF}$ and $R_\mathrm{PTF}$ from this study as a proxy for $g'$ and $r'$, respectively). Figure~\ref{fig:gR_evolution} also shows the color evolution of 35 normal type Ia SNe and 9 underluminous, 91bg-like type Ia SNe from \citet{Folatelli13}. Normal type Ia SNe are defined as those spectroscopically classified as normal by both the Supernova Identification package (SNID; \citealt{Blondin07}) and via the method developed in \citet{Wang09}. 91bg-like SNe are defined as those classified as 91bg-like by SNID. 

The light curves in Figure~\ref{fig:gR_evolution} have been normalized to the time of $B$-band maximum. For 3 (2) 02cx-like SNe: SN~2007qd,  PS1-12bwh, and \fnm\ (SN~2009ku, SN~2015H) the time of $g'$-band ($r'$-band) maximum is used as a proxy because $B$-band observations are not available. Galactic reddening corrections have been applied to all light curves in Figure~\ref{fig:gR_evolution} using the \citet{Schlafly11} updates to the \citet{Schlegel98} dust maps. The following host-galaxy reddening corrections have also been applied $E(B - V) =  0.09$ mag \citep{Phillips07}, $0.50$ mag \citep{Stritzinger14}, $0.20$ mag \citep{Magee17}, and $0.07$ mag \citep{Stritzinger15} for SN~2005hk, SN~2010ae, PS1-12bwh, and SN~2012Z, respectively. The remaining 5 02cx-like SNe do not show evidence of \ion{Na}{1}~D absorption at the redshift of the host galaxy \citep{Foley09,McClelland10,Narayan11,Magee16}, and we therefore make no corrections for host-galaxy reddening. Host-galaxy reddening corrections are not applied to the normal and 91bg-like type Ia SNe as they are not available. As a result there is likely some excess scatter in the curves traced by both the normal and 91bg-like Ia SNe samples in Figure~\ref{fig:gR_evolution}.

The relatively tight locus for 02cx-like color evolution, shown in the left panel of Figure~\ref{fig:gR_evolution}, reveals a previously unknown characteristic of the class. At peak, 02cx-like SNe are very blue similar to normal type Ia SNe (see the left box plot in Figure~\ref{fig:gR_evolution}). While the blue color of 02cx-like SNe near maximum light has clearly been established (e.g., \citealt{Foley13}), the tight scatter, $\sim$0.2 mag, in $g' - r'$ at all epochs in the first $\sim$20 d after peak suggests a common evolution. 

Similar to the correlation between ejecta velocity and luminosity first noted by \citet{McClelland10}, SN~2009ku stands out as a clear outlier from the majority of the 02cx-like class. To bring SN~2009ku in line with the rest of the 02cx-like sample would require a host-galaxy reddening of $E(B - V) \approx 0.35 \; \mathrm{mag}$, which would, in turn, make SN~2009ku $\sim$1.2 mag brighter in the $g'$-band. In this scenario, SN~2009ku would be as luminous, or more luminous, than many normal type Ia SNe. As the sample of 02cx-like SNe grows, the similarity of SN~2009ku to the rest of the class should be closely monitored to determine if SN~2009ku belongs to a new subclass of low-velocity SNe Ia, separate from the other 02cx-like objects.

In addition to following a nearly uniform color curve, 02cx-like SNe exhibit unique $g' - r'$ evolution when compared to normal and 91bg-like type Ia SNe. The right panel of Figure~\ref{fig:gR_evolution} shows the change in $g' - r'$ color relative to $g' - r'$ at $T_{B,\mathrm{max}}$. This normalization enables a measurement of the color evolution that is independent of line-of-sight extinction. It shows that 02cx-like and 91bg-like SNe become significantly redder in the first $\sim$10 d after maximum, while normal type Ia SNe exhibit nearly constant $g' - r'$ color in the same timeframe. To measure this difference we define the change in $g' - r'$ color between maximum and $+10$ d, $\Delta(g' - r')_{10}$. The right inset in Figure~\ref{fig:gR_evolution} shows that 91bg-like and 02cx-like SNe have similar $\Delta(g' - r')_{10}$ values, while normal type Ia SNe have smaller $\Delta(g' - r')_{10}$ values. This, taken in combination with the $g' - r'$ color at peak, provides an empirical method for selecting 02cx-like SNe. At the time of maximum 02cx-like SNe are blue, like normal SNe Ia and unlike 91bg-like SNe, yet they exhibit large $\Delta(g' - r')_{10}$ values, similar to 91bg-like SNe and unlike normal SNe Ia.

The $g' - r'$  color evolution of 02cx-like SNe has important implications for future surveys, such as ZTF and LSST. Spectroscopic completeness will be impossible for these surveys, but it will be possible to identify 02cx-like SNe from their $g' - r'$ color evolution alone. Using our sample of 9 02cx-like SNe, 9 91bg-like SNe, and 35 normal SNe Ia, we summarize the number of each that would be selected following hard cuts on $(g' - r')_{\mathrm{B,max}}$ and $\Delta(g' - r')_{10}$ in Table~\ref{tbl:color_selection}. While we generally advocate against hard cuts for target selection or classification (e.g., \citealt{Miller12}), they illustrate the separation of the 02cx-like class in this case.

{\renewcommand{\arraystretch}{1.15}
\begin{deluxetable}{ccccc}
\tabletypesize{\small}
\tablewidth{3.3in}
\setlength{\tabcolsep}{460pt}
\tablecolumns{5}
\tablecaption{Color Selection Results for 02cx-like SNe \label{tbl:color_selection}}
\tablehead{ \colhead{} & 
            \multicolumn{4}{c}{$(g' - r')_{\mathrm{B,max}}$, $\Delta(g' - r')_{10}$} \\
            \cline{2-5}
            \colhead{SN type} & 
            \colhead{0.0, 0.5\tablenotemark{a}} & 
            \colhead{0.1, 0.4\tablenotemark{a}} & 
            \colhead{0.15, 0.4\tablenotemark{a}} & 
            \colhead{0.15, 0.3\tablenotemark{a}} 
            }
\startdata
Normal Ia & 0/35 & 2/35 & 2/35 & 4/35 \\
91bg-like & 0/9 & 2/9 & 3/9 & 3/9 \\
02cx, obs.\tablenotemark{b} & 3/8 & 4/8 & 5/8 & 5/8\\
\hline
02cx, dered.\tablenotemark{c} & 4/8 & 6/8 & 6/8 & 7/8
\enddata
\tablecomments{Sources bluer than the $(g' - r')_{\mathrm{B,max}}$ cut and with a decline greater than the $\Delta(g' - r')_{10}$ cut are selected as candidate 02cx-like SNe. Our sample includes 9 02cx-like SNe, but SN~2015H is excluded due to a lack of $(g' - r')_{\mathrm{B,max}}$ measurement. SN~2009ku, which has $(g' - r')_{\mathrm{B,max}} \approx 0.34 \; \mathrm{mag}$, is the only 02cx-like SNe that is not selected by any of the above cuts.}
\tablenotetext{a}{Respective cuts on $(g' - r')_{\mathrm{B,max}}$ and $\Delta(g' - r')_{10}$, in mag.}
\tablenotetext{b}{Recovered 02cx-like SNe when host-galaxy-reddening corrections are \textit{not} applied.}
\tablenotetext{c}{Recovered 02cx-like SNe following host-galaxy reddening-corrections, as shown in Figure~\ref{fig:gR_evolution}. }
\end{deluxetable}
}

Table~\ref{tbl:color_selection} shows that 02cx-like SNe are readily separated based on their $g' - r'$ evolution. For instance, adopting cuts of 0.0 and 0.5 (first column of Table~\ref{tbl:color_selection}) selects a pure sample of 02cx-like SNe, though less than half of the sample is recovered. Relaxing the cuts to 0.15 and 0.4 (third column) recovers more than half of the 02cx-like candidates, while still severely restricting the number of false positives.

Using the relative rate of 02cx-like SNe to normal type Ia SNe, 25\% (see \S\ref{sec:rates} below), and the relative rate of 91bg-like SNe to normal type Ia SNe, 19\% \citep{Li11}, we can estimate how many SNe would be selected by these cuts in a volume-limited sample. In a fixed volume with 100 normal SNe Ia we expect to find 19 91bg-like SNe and 25 02cx-like SNe. If we apply cuts of 0.15 and 0.4 on $(g' - r')_{\mathrm{B,max}}$ and $\Delta(g' - r')_{10}$, respectively, we would select $100 \times 2/35 \approx 6$ normal type Ia SNe, $19 \times 3/9 \approx 6$ 91bg-like SNe, and $25 \times 5/8 \approx 16$ 02cx-like SNe. This corresponds to a completeness of $\sim$0.63 and a precision of $\sim$0.56. We caution that this example relies on uncertain rates (\S\ref{sec:rates}), and the assumption that our sample of SNe is representative of what would be found in a volume-limited survey. Starting next year ZTF will significantly reduce the uncertainties on both these assumptions. Nevertheless, this selection function for 02cx-like SNe will enable the efficient use of resources in the near future when transients are plentiful and follow-up scarce.

\subsection{The Relative Rate of 02cx-like SNe}\label{sec:rates}

The controlled nature of the \textit{Color Me Intrigued} experiment enables a unique estimate of the relative rate of 02cx-like SNe compared to type Ia SNe. Assuming 02cx-like SNe peak at $M \approx -15$~mag, then iPTF, which has a detection limit of $R_\mathrm{PTF} \leq 20.5$~mag \citep{law09}, can detect these sources to a distance modulus $\mu = 35.5$~mag, corresponding to redshift $z \leq 0.03$. Below this redshift limit, there were 4 SNe Ia and 1 02cx-like SN, \fnm, discovered during the course of the \textit{Color Me Intrigued} experiment. Under the assumption of observational completeness, we therefore find that the relative rate of 02cx-like SNe to type Ia SNe, $r_{\mathrm{02cx}/\mathrm{Ia}}$, is 1/4 = 25\%.

Prior to estimating the uncertainty on $r_{\mathrm{02cx}/\mathrm{Ia}}$, we caution that our assumption of observational completeness for 02cx-like SNe during the \textit{Color Me Intrigued} experiment is likely overly optimistic. First, the luminosity function (LF) of 02cx-like SNe is poorly constrained. If the LF is heavily weighted towards extremely faint 08ha-like SNe, iPTF would have missed any of these beyond $z \approx 0.02$. Second, despite the high-cadence observations during \textit{Color Me Intrigued}, the $\sim$week-long gaps in monitoring around full moon could have resulted in additional 02cx-like SNe that were missed. Third, our machine learning candidate identification software, \texttt{realbogus}, is limited by the flux contrast between the host galaxy and the SN. iPTF is complete to transients that are $>10 \times$ brighter than the underlying host surface brightness, but when the contrast drops to $0.7 \times$ ($0.2 \times$) the host galaxy surface brightness the completeness drops to $\sim$50\% ($\sim$2\%), resulting in missed transients (Frohmaier et al.\ 2017, ApJ submitted). These concerns should not affect type Ia SNe, which are more luminous and long lived, meaning our $z \leq 0.03$ sample is potentially biased against 02cx-like SNe. Thus, the rate estimates included here are likely underestimates. Nevertheless, we proceed under the assumption that iPTF detected all $z \leq 0.03$ 02cx-like SNe during the \textit{Color Me Intrigued} experiment.

Following the analysis presented in \citet{White15}, we can calculate confidence intervals for the relative rate of 02cx-like SNe to type Ia SNe. Given an outcome with probability $p$, the chances of getting $k$ successes in a sample of $n$ trials is given by the binomial distribution probability mass function:
\begin{equation}
    Pr(k;n,p) = \binom{n}{k} p^k (1-p)^{n-k}.
\end{equation}
Assuming a uniform prior [0,1], the probability of $p$ being less than some fiducial value $p_0$ given an observed fraction $k/n$ is
\begin{equation}
    Pr(p < p_0;n,k) = \frac{\int_0^{p_0} Pr(k;n,p) dp}{\int_0^1 Pr(k;n,p) dp}.
\end{equation}
From here, it follows that the probability that the relative rate $r_{\mathrm{02cx}/\mathrm{Ia}}$ is less than $r_0$ given $N_\mathrm{02cx}$ 02cx-like SNe and $N_\mathrm{Ia}$ type Ia SNe is
\begin{equation}
    Pr^\mathrm{rel}_{N_\mathrm{02cx},N_\mathrm{Ia}}(r_{\mathrm{02cx}/\mathrm{Ia}} < r_0) = Pr(p < \frac{r_0}{1 + r_0}; N_\mathrm{SN}, N_\mathrm{02cx}),
\end{equation}
where $N_\mathrm{SN} = N_\mathrm{Ia} + N_\mathrm{02cx}$.

Considering SNe within the 02cx-like volume-limited sample (i.e., $z \leq 0.03$), we have $N_\mathrm{Ia} = 4$ and $N_\mathrm{02cx} = 1$. We find $Pr^\mathrm{rel}_{1,4}(r < 100\%) = 95\%$ and $Pr^\mathrm{rel}_{1,4}(r < 6.5\%) = 5\%$, which corresponds to a 90\% confidence interval for the relative rate of $25^{+75}_{-18.5}\%$. In this case, small number statistics result in an estimate of the relative rate that is not particularly constraining. Nevertheless it is consistent with previous estimates including $\sim$5\% from \citet{Li11}, 5.6\% from \citet{White15}, and 31\% from \citet{Foley13}. 
While the methodologies differ, there is general agreement within the (large) uncertainties. Additionally, \citeauthor{Li11} note that their rate is likely underestimated if the luminosity function of 02cx-like SNe extends significantly fainter than SN~2005hk. Given that there are now many known 02cx-like SNe fainter than SN~2005hk (e.g., SNe~2008ha,~2007qd,~2010ae,~\fnm), it stands to reason that the true rate is likely higher than the estimate in \citeauthor{Li11} After limiting the survey volume to very nearby SNe, \citet{Foley13} and \citet{White15} apply correction factors of 2 and 1, respectively, to their relative rate measurements. The factor of 2 adopted in \citeauthor{Foley13} is highly uncertain, given the poorly-constrained LF of 02cx-like SNe and the many heterogeneous surveys used to define their sample of 02cx-like SNe. Meanwhile, the assumption that PTF was spectroscopically complete for slow-speed SNe, which is adopted in \citeauthor{White15}, is likely overly optimistic. The true value of $r^\mathrm{rel}$ is likely between 0.05 and 0.3, and future surveys with large volumetric survey speeds \citep{Bellm16a}, such as ZTF, are needed to significantly reduce the uncertainty on this measurement.

\section{Summary and Conclusions}

Maximizing information content while maintaining a large discovery rate is a challenge for modern wide-field time-domain surveys. This challenge will only be exacerbated in the coming years as extremely large field-of-view (ZTF) and large aperture (LSST) surveys come online. The looming orders-of-magnitude increase in discovered transients will overwhelm existing follow-up facilities and generate a ``follow-up problem.'' Ultimately this means survey telescopes will provide the sole observations of a majority of transients discovered in the coming decade. This near-future reality necessitates the immediate development of photometric-only methods for studying SNe and other transients.

To address the ``follow-up problem'' in the context of ZTF, we recently completed the \textit{Color Me Intrigued} experiment during the final semester of iPTF. \textit{Color Me Intrigued} searched for transients simultaneously in the $g_\mathrm{PTF}$ and $R_\mathrm{PTF}$ filters, marking the first time this was done in PTF/iPTF. \textit{Color Me Intrigued} addressed the ``follow-up problem'' in two critical ways: (i) the color information allowed us to identify extreme transients \textit{at the epoch of discovery} by winnowing those with typical colors, ensuring only rare objects receive time-critical follow-up observations, and (ii) the strategy guaranteed that all newly discovered transients had color information at all epochs, which is critical for classifying sources without follow-up observations (e.g., \citealt{Poznanski02}). As all PTF/iPTF surveys require two observations of the same field on a given night to reject asteroids as non-explosive transients, \textit{Color Me Intrigued} provided a significant addition of information without a loss of survey area. As a precursor for LSST, we further employed a mixed cadence strategy whereby we cycled through 3 groups of fields each of which was observed with 1-d cadence for 1 lunation, and 3-d cadence at all other times. This mixed-cadence strategy was adopted to achieve a uniform depth across all fields at the end of the survey, while also monitoring variability on short and long timescales.

During the course of \textit{Color Me Intrigued}, we discovered \fnm, a new member of the 02cx-like subclass of type Ia SNe. \fnm\ peaked at $M_{g_\mathrm{PTF}} = -15.09 \pm 0.17 \; \mathrm{mag}$ and declined by $\sim$1.2 mag in 13 d. The spectra of \fnm\ were emblematic of the 02cx-like class, including the following properties: (i) very low velocity ejecta ($v_\mathrm{ej} \approx {3000} \; \mathrm{km \; s}^{-1}$) , (ii) strong absorption from intermediate mass and Fe-group elements, and (iii) a not-fully nebular appearance several months after peak luminosity. Based on its photometric and spectroscopic evolution,  \fnm\ is an unambiguous member of the 02cx-like class.

We additionally compared \fnm\ to other 02cx-like SNe, and find that it is among the least luminous members of the class. \fnm\ is the second faintest known SN Ia, after SN~2008ha, which peaked at $M_{g'} = -14.01 \pm 0.14 \; \mathrm{mag}$ \citep{Stritzinger14}.\footnote{Small changes in the reddening corrections or distance modulii to SN~2007qd or SN~2010ae could shuffle the order of the $2^\mathrm{nd}$, $3^\mathrm{rd}$, and $4^\mathrm{th}$ least luminous SNe Ia.} The post-peak spectra of \fnm\ exhibit a striking resemblance to those of SNe~2007qd and 2010ae, with similar velocities and chemical compositions. These two SNe also peaked at $M \approx -15 \; \mathrm{mag}$, and they exhibit a similar light curve evolution to \fnm. The nearly identical evolution of these 3 SNe suggests a common origin.

Many studies have looked for correlated properties (luminosity, ejecta velocity, decline rate, etc.) to see if 02cx-like SNe can be described as a single-parameter family, similar to normal type Ia SNe. We update the previous work of \citet{White15} and find that, at best, there is a weak correlation with large scatter between absolute magnitude and $\Delta{m}_{15}$ in both the $g'$- and $r'$-bands. We also examine the $g' - r'$ color evolution of 02cx-like SNe, and compare it to normal and 91bg-like type Ia SNe. We find that 02cx-like SNe 
exhibit unique color evolution: blue colors at peak with large $\Delta(g' - r')_{10}$ values. While we do not have a physical explanation for this behavior, this empirical result can be used as a selection function for identifying 02cx-like SNe. We show that simple cuts on $(g' - r')_{B,\mathrm{max}}$ and $\Delta(g' - r')_{10}$ selects 02cx-like SNe with high completeness and precision. While limited by small number statistics, we nevertheless measure the relative rate of 02cx-like SNe to normal SNe Ia and find $r_{N_{02cx}/N_{Ia}} = 25^{+75}_{-18.5}\%$. This measurement is consistent with other estimates in the literature \citep{Li11,Foley13,White15}.

In advance of ZTF, the \textit{Color Me Intrigued} experiment has demonstrated that nightly observations in different filters can efficiently discover transients.
In fact, the general success of this experiment has led to the decision to conduct the $3\pi$ ZTF public survey\footnote{ZTF is a public-private partnership with 40\% of the telescope time dedicated to a public survey. The public survey will monitor the full sky observable from Palomar Observatory with a 3-d cadence.} with near-simultaneous $g_\mathrm{ZTF}$ and $r_\mathrm{ZTF}$ observations. The experiment has further shown the power of closely coupling efficient follow-up resources to survey telescopes. In particular, the SEDm correctly identified \fnm\ as a low-velocity SN, despite its low spectral resolution ($R \approx 100$). Indeed, the SEDm model provides one potential path toward reducing the ``follow-up problem'' for LSST: building low-resolution spectrographs for 4-m class telescopes would enable efficient follow up for LSST transients with $r' \lesssim 22 \; \mathrm{mag}$. 

We close with a recommendation for future time-domain surveys. The search for astrophysical transients provides a fast moving target where new phenomena are regularly uncovered. These new discoveries often require new observational strategies to efficiently increase the sample size of these rarities. Due to its unique aperture and survey capabilities, early observations from LSST will likely uncover new phenomena.\footnote{To highlight an example from PTF, commisioning observations resulted in the discovery of 3 super-luminous SNe \citep{quimby11}. The \citeauthor{quimby11} result established a new class of stellar explosion, while also explaining the nature of SCP 06F6 \citep{Barbary09}, the most mysterious optical transient known at that time.} If the LSST observational strategy is fixed without flexibility from the start of the survey, then the transient discoveries in year 2 will look like those from year 1, while year 3 will look like year 2, and so on. This scenario is detrimental for the exploration of explosive systems. By emphasizing a change in cadence at periodic intervals, iPTF was able to specifically target rare sources identified by PTF (e.g., the extensive use of 1-d cadence to find more young SNe similar to PTF~11kly/SN~2011fe; \citealt{Nugent11,Bloom12a,Cao16}), while also enabling new methods of exploration, such as \textit{Color Me Intrigued}. Thus, we advocate for the adoption of some measure of flexibility in the observational strategy of LSST. Even if minor, this flexibility may prove crucial to better understanding the nature of unusual sources discovered in the early stages of LSST.

\textit{ Facilities:} 
\facility{PO:1.2m}, \facility{PO:1.5m (SEDm)}, \facility{Hale (DBSP)}, \facility{NOT (ALFOSC)}, \facility{Keck:I (LRIS)}

\textcopyright 2017. All rights reserved.

\acknowledgments 

We thank K.~Shen for a useful discussion on the color evolution of Type Ia SNe. R.~Amanullah pointed us to several useful papers on the color evolution of Type Ia SNe, for which we are grateful. Finally, we thank J.~Nordin for examining the spectra of type Ia SNe found during \textit{Color Me Intrigued}.

The Intermediate Palomar Transient Factory project is a scientific collaboration among the California Institute of Technology, Los Alamos National Laboratory, the University of Wisconsin, Milwaukee, the Oskar Klein Center, the Weizmann Institute of Science, the TANGO Program of the University System of Taiwan, and the Kavli Institute for the Physics and Mathematics of the Universe. This work was supported by the GROWTH project funded by the National Science Foundation under Grant No 1545949. Part of this research was carried out at the Jet Propulsion Laboratory, California Institute of Technology, under a contract with NASA.

Some of the data presented here were obtained in part with ALFOSC, which is provided by the Instituto de Astrofisica de Andalucia (IAA) under a joint agreement with the University of Copenhagen and NOTSA.

Some of the data presented herein were obtained at the W.M. Keck Observatory, which is operated as a scientific partnership among the California Institute of Technology, the University of California and NASA. The Observatory was made possible by the generous financial support of the W.M. Keck Foundation. 

The authors wish to recognize and acknowledge the very significant cultural role and reverence that the summit of Mauna Kea has always had within the indigenous Hawaiian community.  We are most fortunate to have the opportunity to conduct observations from this mountain.

This research has made use of the NASA/IPAC Extragalactic Database (NED), which is operated by the Jet Propulsion Laboratory, California Institute of Technology, under contract with NASA.

\end{document}